\documentclass[acmsmall]{acmart}

\setcopyright{cc}
\setcctype{by}
\acmDOI{10.1145/3832166}
\acmYear{2026}
\acmJournal{PACMSE}
\acmVolume{3}
\acmNumber{ISSTA}
\acmArticle{ISSTA075}
\acmMonth{10}
\acmSubmissionID{issta26main-p725-p}
\received{2026-01-30}
\received[accepted]{2026-06-25}

\usepackage{xspace}
\usepackage{xcolor}
\usepackage{graphicx}
\usepackage[normalem]{ulem} 
\usepackage{amsmath}
\usepackage{enumitem}
\usepackage{needspace}

\usepackage{hyperref}
\hypersetup{linkcolor=black,citecolor=black,anchorcolor=black,filecolor=black,menucolor=black,runcolor=black,urlcolor=black,hidelinks}
\usepackage{breakurl}

\usepackage{booktabs}
\usepackage{multirow}
\usepackage{makecell}
\usepackage{ragged2e}
\usepackage{colortbl}
\usepackage{threeparttable}
\arrayrulecolor{black}

\usepackage{caption}
\usepackage{subcaption}
\usepackage{tcolorbox}
\usepackage{wrapfig} 

\usepackage{listings}

\usepackage{tikz}
\usetikzlibrary{calc}
\usetikzlibrary{shapes.geometric}
\usetikzlibrary{decorations.pathreplacing}
\usetikzlibrary{positioning}

\usepackage{datetime} 

\makeatletter
\newcommand{\DefMacro}{\@ifstar\@DefMacroAllowRedefine\@DefMacro}
\newcommand{\@DefMacro}[2]{\expandafter\newcommand\csname rmk-#1\endcsname{#2}}
\newcommand{\@DefMacroAllowRedefine}[2]{\expandafter\providecommand\csname rmk-#1\endcsname{} \expandafter\renewcommand\csname rmk-#1\endcsname{#2}}
\makeatother
\newcommand{\UseMacro}[1]{\csname rmk-#1\endcsname}

\newcommand{\revise}[1]{#1}
\newcommand{\new}[1]{#1}

\newcommand{\EditRm}[1]{}

\newcommand{\EditMvRm}[1]{}

\newcommand{\XSpace}[1]{}
\newcommand{\XComment}[1]{}

\newcommand{\MyPara}[1]{\noindent\textbf{#1}.}

\newcommand{\Code}[1]{{\ifmmode{\mathtt{#1}}\else$\mathtt{#1}$\fi}}
\newcommand{\CodeIn}[1]{{\ifmmode{\mathtt{#1}}\else$\mathtt{#1}$\fi}}

\newcolumntype{R}[1]{>{\RaggedLeft\arraybackslash}p{#1}}
\newcolumntype{L}[1]{>{\RaggedRight\arraybackslash}p{#1}}

\definecolor{gray}{RGB}{211,211,211}
\newcommand{\jbasicstyle}{\small\sffamily} 

\newcommand{\jnumberstyle}{\scriptsize}

\lstdefinelanguage{pseudo}
{
  morekeywords={},
  keywordstyle=\bfseries,
  lineskip=-0.1em,
  numbers=left, 
  numberstyle=\jnumberstyle,
  numbersep=4pt,
  basicstyle=\jbasicstyle,
  breaklines=true,
  breakautoindent=true,
  tabsize=2,
  columns=fullflexible,
  morecomment=*[l][\textsl]{//},
  mathescape=true,
  xleftmargin=10pt,
}

\lstdefinelanguage{todo-comment}
{
  morekeywords={},
  keywordstyle=\bfseries,
  lineskip=-0.1em,
  numbers=none,
  basicstyle=\jbasicstyle,
  breaklines=true,
  breakautoindent=true,
  tabsize=2,
  columns=fullflexible,
  morecomment=*[l][\textsl]{//},
  mathescape=true,
  xleftmargin=-10pt,
}

\lstdefinelanguage{java-pretty}
{
  language=java,
  numbers=left,
  basicstyle=\scriptsize\ttfamily,
  numberstyle=\scriptsize,
  breaklines=true,
  columns=fullflexible,
  xleftmargin=16pt,
  showstringspaces=false,
}

\newcommand*{\conclusionbox}[1]{%
    \vspace{1mm}
    \noindent
    \fcolorbox{black}{gray!10}{%
        \parbox[b]{0.96\columnwidth}{%
            \emph{#1}
        }
    }
}

\newcommand{\Tool}{\textsc{MLEModernizer}\xspace}
\newcommand{\ToolURL}{\url{https://github.com/Bihui-Jin/MLEModernizer}}
\newcommand{\Title}{Automated Modernization of Machine Learning Engineering Notebooks for Reproducibility}

\newcommand{\eg}{e.g.,\xspace}

\newcommand{\reproducibility}{reproducibility\xspace}
\newcommand{\Reproducibility}{Reproducibility\xspace}
\newcommand{\nonreproducibility}{non-reproducibility\xspace}
\newcommand{\reproducible}{reproducible\xspace}
\newcommand{\nonreproducible}{non-reproducible\xspace}
\newcommand{\reproduce}{reproduce\xspace}

\newcommand{\backporting}{backporting\xspace}
\newcommand{\nb}{notebook\xspace}
\newcommand{\nbs}{notebooks\xspace}
\newcommand{\Nb}{Notebook\xspace}
\newcommand{\Nbs}{Notebooks\xspace}
\newcommand{\envbackporting}{environment backporting\xspace}
\newcommand{\Envbackporting}{Environment backporting\xspace}
\newcommand{\score}{score\xspace}
\newcommand{\scores}{scores\xspace}
\newcommand{\targetscore}{target score\xspace} 
\newcommand{\targetscores}{target scores\xspace}
\newcommand{\reproducedscore}{reproduced score\xspace}
\newcommand{\reproducedscores}{reproduced scores\xspace}
\newcommand{\scoredeviation}{score deviation\xspace}
\newcommand{\errorrepair}{error-repair\xspace}
\newcommand{\ErrorRepair}{Error-Repair\xspace}
\newcommand{\runtimereduction}{runtime-reduction\xspace}
\newcommand{\RuntimeReduction}{Runtime-Reduction\xspace}
\newcommand{\scorecalibration}{score-calibration\xspace}
\newcommand{\ScoreCalibration}{Score-Calibration\xspace}

\newcommand{\dockerimages}{Docker images\xspace}
\newcommand{\dockercontainer}{Docker container\xspace}
\newcommand{\dockercontainers}{Docker containers\xspace}
\newcommand{\container}{container\xspace}

\newcommand{\bsl}{Baseline\xspace}
\newcommand{\dg}{Backporting\xspace}

\newcommand{\rep}{Reproducible\xspace}
\newcommand{\srep}{reproducible\xspace}
\newcommand{\nrep}{Non-reproducible\xspace}

\newcommand{\xEFNR}{Error-Free Non-Reproducible\xspace}
\newcommand{\xEFR}{Error-Free Reproducible\xspace}
\newcommand{\xENR}{Error Non-Reproducible\xspace}
\newcommand{\xER}{Error Reproducible\xspace}

\newcommand{\Jupyternbs}{Jupyter notebooks\xspace}
\newcommand{\Kaggle}{Kaggle\xspace}
\newcommand{\MLEBench}{MLE-Bench\xspace}

\newcommand{\csv}{CSV\xspace}

\DefMacro{base-llm}{GPT-5.2\xspace}

\newcommand{\aTargetScore}{\CodeIn{s_t}\xspace}
\newcommand{\aReproducedScore}{\CodeIn{s_r}\xspace}

\newcommand{\aThreshold}{\CodeIn{\tau}\xspace}
\newcommand{\aMaxIterations}{\CodeIn{N}\xspace}

\DefMacro{TH-error-free-reproducible}{EF-R\xspace}
\DefMacro{TH-error-free-non-reproducible}{EF-NR\xspace}
\DefMacro{TH-error-reproducible}{E-R\xspace}
\DefMacro{TH-error-non-reproducible}{E-NR\xspace}
\DefMacro{TH-statistical-t-test}{Statistical $t$-test\xspace}

\newcommand{\PerfThreshold}{10\%\xspace}
\newcommand{\BaselineReproducibility}{\revise{26\%}\xspace}
\newcommand{\BackportReproducibility}{\revise{12\%}\xspace}
\newcommand{\NumNotebookAll}{\revise{12,106}\xspace}
\newcommand{\NumNotebookSubject}{\revise{8,210}\xspace}
\newcommand{\NumCompetition}{\revise{75}\xspace}
\newcommand{\FileReproducibleNumNotebook}{\revise{3,292}\xspace}
\newcommand{\FileReproducibility}{\revise{40.1\%}\xspace}
\newcommand{\FileErrorReproducibleNumNotebook}{\revise{895}\xspace}
\newcommand{\FileErrorReproducibility}{\revise{10.9\%}\xspace}
\newcommand{\AvgFixIterations}{\revise{10.1\xspace}}
\newcommand{\AvgCostPerNotebook}{\revise{0.65\xspace}}

\begin{document}

\title{\Title}


\author{Bihui Jin}
\orcid{0009-0009-0011-1134}
\affiliation{
  \institution{University of Waterloo}
  \city{Waterloo}
  \country{Canada}
}
\email{bihui.jin@uwaterloo.ca}

\author{Kaiyuan Wang}
\orcid{0000-0002-1790-0721}
\affiliation{
  \institution{Google Inc.}
  \city{Mountain View}
  \country{USA}
}
\email{kaiyuanw@google.com}

\author{Pengyu Nie}
\orcid{0000-0003-1529-3216}
\affiliation{
  \institution{University of Waterloo}
  \city{Waterloo}
  \country{Canada}
}
\email{pynie@uwaterloo.ca}

\begin{abstract}

Interactive computational notebooks (e.g., \Jupyternbs) are widely used in machine learning engineering (MLE) to program and share end-to-end pipelines, from data preparation to model training and evaluation.
However, \emph{environmental erosion}---the rapid evolution of hardware and software ecosystems for machine learning---has rendered many published MLE \nbs \nonreproducible in contemporary environments, hindering code reuse and scientific progress.
To quantify this gap, we study \NumNotebookAll \nbs selected from \NumCompetition popular \Kaggle competitions: only \BaselineReproducibility remain \reproducible today.
Crucially, we find that \envbackporting, i.e., downgrading dependencies to match the submission time, does not improve \reproducibility (decreased to \BackportReproducibility) but rather introduces additional failure modes.

To address environmental erosion, we design and implement \Tool, an LLM-driven agentic framework that treats the contemporary environment as a fixed constraint and modernizes notebook code to restore \reproducibility.
\Tool iteratively executes notebooks, collects execution feedback, and applies three types of targeted fixes: \errorrepair, \runtimereduction, and \scorecalibration.
Evaluated on \NumNotebookSubject \nbs that are \nonreproducible under the baseline environment, \Tool makes \FileReproducibleNumNotebook (\FileReproducibility, GPT-5.2) and 3,683 (44.9\%, GPT-OSS-120b) \nbs \reproducible. 
\Tool \new{presents a best-effort automated recovery and modernization technique that can improve \reproducibility for a subset of notebooks.
Practitioners can leverage \Tool to validate, reuse, and maintain MLE artifacts as the hardware and software ecosystems continue to evolve.} 

\end{abstract}

%

\ccsdesc[500]{Software and its engineering~Software evolution}
\ccsdesc[500]{Software and its engineering~Software maintenance tools}
\ccsdesc[300]{Computing methodologies~Machine learning}
%

\keywords{Machine learning engineering, code modernization, Jupyter notebooks}

\maketitle


\section{Introduction}
\label{sec:intro}

Interactive computational notebooks, such as \Jupyternbs~\cite{KluyverETAL16Jupyter}, are widely adopted in machine learning and data science~\cite{WangETAL21DocumentationNotebook,YinETAL23CodeGenNotebooks,GhahfarokhiETAL24DistilKaggle,PimentelETAL19JupyterRepro},
thanks to their combination of interactive code execution and rich visualizations~\cite{RuleETAL18Notebooks}.
In machine learning engineering (MLE), Python \Jupyternbs (hereafter referred to as \emph{\nbs}) serve as the primary code artifacts for documenting and sharing end-to-end machine learning pipelines, from data preparation to model training and evaluation~\cite{WangETAL21DocumentationNotebook,PimentelETAL19JupyterRepro}.
Many machine learning research prototypes are shared in this format~\cite{PimentelETAL19JupyterRepro,GhahfarokhiETAL24DistilKaggle}.

\Reproducibility is a crucial property in MLE to support trustworthy scientific progress in academia, and to facilitate code reuse in industry.
Nonetheless, MLE \nbs increasingly suffer from the \emph{environmental erosion} problem: as hardware and software ecosystems rapidly evolve, re-executing older notebooks in a current environment often fails or yields different results, and reconstructing the exact historical environment is rarely feasible~\cite{PimentelETAL19JupyterRepro,PineauETAL21ReproducibilityNeurIPS,GundersenKjensmo18ReproAI}.
This motivates a practical question for both research and reuse: can we reproduce the reported \scores of previously published \nbs on the latest hardware and software stack?

To assess this, we conduct a motivation study on \NumNotebookAll \nbs selected from \NumCompetition popular \Kaggle competitions~\cite{Chan24MLEBench}.
We re-execute the \nbs using a recent \Kaggle \dockercontainer (aligned with how \Kaggle executes submissions), classifying \reproducibility via repeated executions and one-sample $t$-tests. 
Our study shows that only \BaselineReproducibility of the \nbs are \reproducible; the \reproducibility gap persists despite a relatively standardized ecosystem~\cite{PimentelETAL19JupyterRepro,PimentelETAL21Julynter}.

The standard industry response of \envbackporting, i.e., downgrading dependencies to match the \nb's submission time, does not close this gap.
Since \Kaggle's \dockerimages older than two years are no longer available, we implement a \backporting algorithm that downgrades dependencies to the submission timestamp.
Surprisingly, only \BackportReproducibility of \nbs remain \reproducible after \backporting, substantially lower than the baseline \reproducibility rate of \BaselineReproducibility.
This result exposes a fundamental flaw in the prevailing paradigm: treating the environment as a variable to be reconstructed and the notebook code as an immutable artifact.
We therefore shift our focus to modernizing the notebook code to restore \reproducibility, so that \nbs are modernized to run and reproduce under the current environment rather than the historical one.
Repair in this setting cannot stop at ``green'' (no errors): an MLE \nb that runs but whose accuracy drops from 95\% to 60\% is not acceptable.
We thus treat \scorecalibration---moving the reproduced score toward the reported \Kaggle score within the \reproducible band---as a first-class repair target alongside \errorrepair and \runtimereduction.

Recent advances in large language models (LLMs) have shown promising coding capabilities, especially for program repair~\cite{BouzeniaETAL25RepairAgent,XieETAL25PReMM} and code evolution tasks, such as adapting deprecated APIs~\cite{WangETAL25DeprecatedAPI,DilharaETAL23PyEvolve}.
In this work, we explore the use of LLMs to \emph{modernize} MLE \nbs so that they can become \reproducible in contemporary environments.
We design and implement \Tool, an LLM-driven agentic framework that iteratively fixes a \nb toward reproducibility.
Each iteration begins with \Tool executing the \nb and recording its score deviation (from the reported \score), runtime, and any error messages during execution.
By analyzing the execution context, \Tool prepares a targeted LLM prompt for one of the three fix types: \errorrepair, \scorecalibration, or \runtimereduction.
Then, \Tool utilizes the LLM to generate a patch at the file level (i.e., the entire \nb), enabling it to identify the root cause of \nonreproducibility and propose a comprehensive solution covering multiple code locations (thus reducing the number of fix iterations needed).

To evaluate \Tool, we apply it to \NumNotebookSubject \nonreproducible \nbs in our dataset, using \UseMacro{base-llm} as the primary base LLM and GPT-OSS-120b as a comparison model.
\Tool successfully modernizes \FileReproducibility--44.9\% of \nbs, making \FileReproducibleNumNotebook notebooks \reproducible with \UseMacro{base-llm} and 3,683 \reproducible with GPT-OSS-120b. 
Notably, \FileErrorReproducibleNumNotebook \nbs (\FileErrorReproducibility) become \reproducible despite residual errors (\xER); these errors do not prevent the \nbs from meeting the score-based criterion, but they may indicate unresolved issues in the accompanying visualization or analysis code.
\Tool is also economically viable: it requires \AvgFixIterations fix iterations on average and costs an average of \$\AvgCostPerNotebook USD per \nb, making it practical for large-scale industrial use.
\revise{We further validate the modernization outcomes by manually inspecting a statistically sampled subset of before-and-after \nb pairs to assess functional equivalence, complemented by a systematic analysis of error types and failure modes. We find that although GPT-OSS-120b is slightly more robust in achieving \reproducibility, GPT-5.2 produces a substantially higher proportion of functionally equivalent modernizations (71.5\% vs. 52.7\%). }

Our work has broad implications for the MLE community.
By enabling the automated modernization of MLE notebooks, \Tool can facilitate the validation and reuse of MLE pipelines even as software and hardware stacks rapidly evolve. \new{
Because exact reproduction may be impossible due to the evolved environment, the best-effort modernization by our approach can serve as a starting point for manual refinement}.
Researchers can use \Tool to \reproduce \new{legacy notebooks, reducing the tedious manual effort, thereby facilitating knowledge transfer in ML research.}
Engineers can use \Tool to \new{help update legacy MLE pipelines and enhance maintainability and \reproducibility, thereby building user trust in their long-term usability}.

\vspace{3pt}
\noindent
The main contributions of this work include:
\begin{itemize}[topsep=3pt,itemsep=1ex,partopsep=0ex,parsep=0ex,leftmargin=*]
\item \MyPara{Motivation Study}
We conduct a \reproducibility study of MLE \nbs mined from popular \Kaggle competitions and show that only a minority remain \reproducible today.
We further demonstrate that \envbackporting does not close the \reproducibility gap.
\item \MyPara{Technique}
We develop \Tool, an LLM-driven agentic framework \new{as an initial attempt to recover and} modernize MLE \nbs for reproducibility in contemporary environments via three targeted fix types: \errorrepair, \runtimereduction, and \scorecalibration.
\item \MyPara{Evaluation}
We evaluate \Tool and find that it successfully modernizes \FileReproducibleNumNotebook (\FileReproducibility) of the \nonreproducible \nbs in our dataset, with a low cost of \$\AvgCostPerNotebook USD per \nb, demonstrating practicality for large-scale use.
\item \MyPara{Dataset}
We curate a dataset of real-world \Kaggle MLE \nbs with execution traces and \reproducibility labels, enabling future research on MLE \nbs.
\end{itemize}

\noindent
Our code and data are publicly available at \ToolURL.

\section{Motivation}
\label{sec:motivation}

\begin{figure}[t]
\centering
\includegraphics[width=\textwidth]{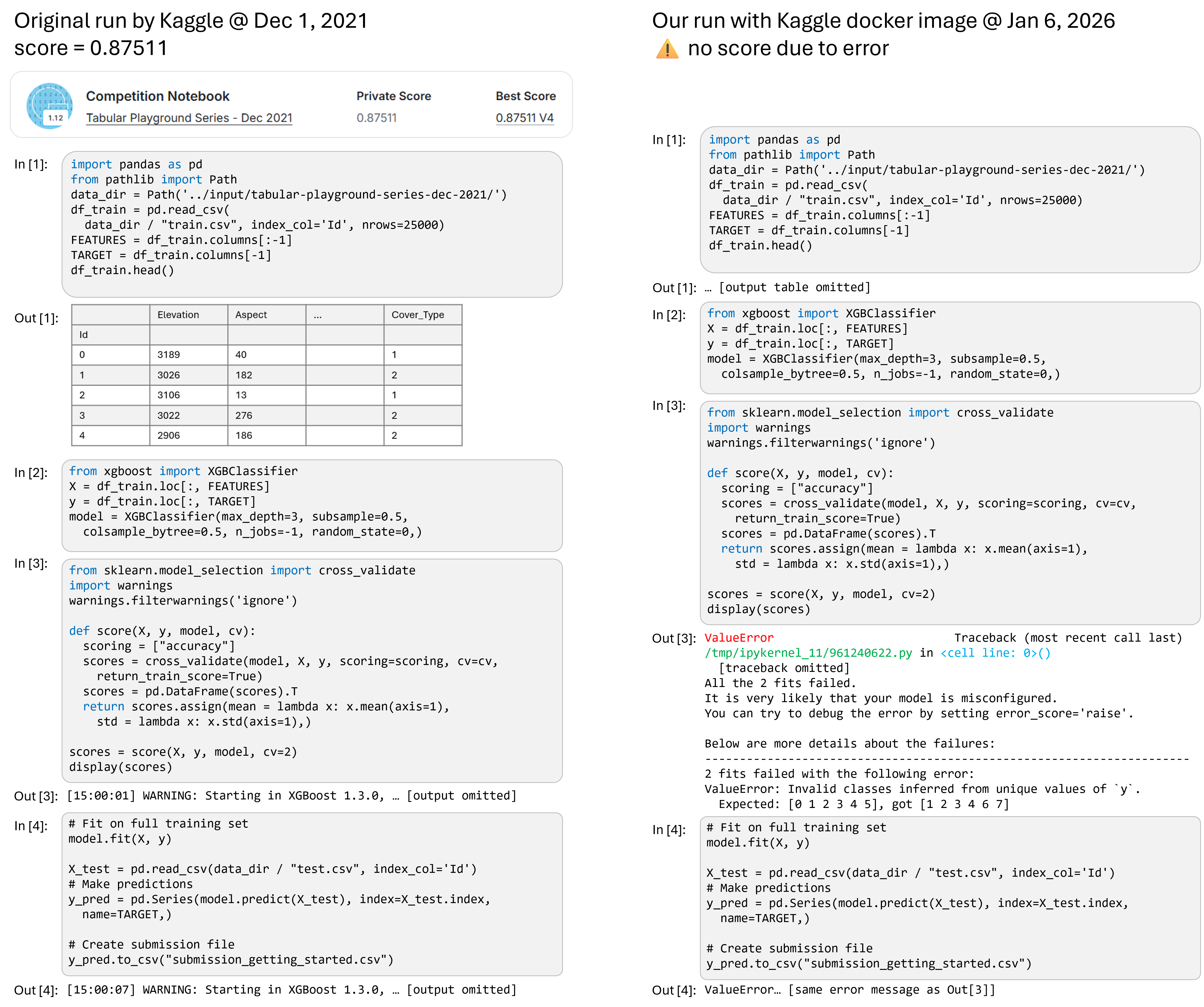}
\vspace{-22pt}
\caption{An example \nb and our attempt to \reproduce its results. 
Left: successful run on \Kaggle; Right: our run failed due to a breaking change in XGBoost.}
\label{fig:motivation-example}
\end{figure}

To demonstrate the \reproducibility challenges in MLE \nbs, we show an example \nb submitted to the ``Tabular Playground Series'' \Kaggle competition\footnote{Link to the \nb: \url{https://www.kaggle.com/code/sugamkhetrapal/tps-dec-2021-1-05-getting-started}; Link to the competition: \url{https://www.kaggle.com/competitions/tabular-playground-series-dec-2021/overview}.} in Figure~\ref{fig:motivation-example}.
\Kaggle is a prominent platform for MLE competitions and code sharing.
The MLE pipeline in this \nb contains four steps (i.e., cells labeled as ``In[x]'' and ``Out[x]''):
(1)~loading the training set;
(2)~configuring an XGBoost classifier model with hyper-parameters;
(3)~cross-validating the model on a subset of the training set, which is useful for hyper-parameter tuning;
(4)~training the model on the full training set, then applying the model on the test set, and finally generating a \csv file with predictions for submission.
The left part of Figure~\ref{fig:motivation-example} shows the successful run on \Kaggle when the \nb was submitted, where the \nb obtained a \score (prediction accuracy) of 0.87511 in the competition.
However, as shown in the right part of Figure~\ref{fig:motivation-example}, our attempt to \reproduce its results failed with errors when training the model.
The errors are caused by a breaking change in XGBoost for target class inference:
the ``Cover\_Type'' labels in the original dataset range from 1 to 7,
but newer versions of XGBoost require labels to be consecutive integers starting from 0.

In \S\ref{sec:motivation:baseline}, we conduct a large-scale study to show that \nonreproducibility (such as that caused by breaking changes in dependency libraries) is a pervasive problem in MLE \nbs.
Then, in \S\ref{sec:motivation:backporting}, we show that \envbackporting, a naive solution that attempts to \reproduce a \nb by downgrading the dependency libraries' versions to match the submission timestamp of the \nb, does not solve the \reproducibility challenges.

\subsection{\Reproducibility of MLE \Nbs}
\label{sec:motivation:baseline}

\subsubsection{Dataset Collection}
\label{sec:motivation:baseline:data}

To systematically study the \reproducibility challenges in MLE \nbs, we construct a large-scale dataset of real-world MLE \nbs from \Kaggle. 
Our dataset collection process involves two main steps: 
(1)~identifying relevant submissions from \Kaggle competitions, and 
(2)~extracting comprehensive metadata and code content for each submission version.

\paragraph{Competition Selection}
We start from \MLEBench~\cite{Chan24MLEBench}, a curated benchmark built from \Kaggle competitions and designed for evaluating end-to-end ML workflows.
\MLEBench provides a standardized offline grading setup, which makes it feasible to execute and grade \nbs at scale when \Kaggle's held-out test sets are not publicly available.
By anchoring our study on the competitions in \MLEBench, we cover a broad range of high-quality MLE tasks including image classification, natural language processing, time series prediction, and tabular data analysis.

\paragraph{Notebook Mining}
\revise{
We leverage \Kaggle's official data dump, Meta Kaggle~\cite{MetaKaggle} and Meta Kaggle Code~\cite{MetaKaggleCode}, to identify all notebook submissions associated with the selected competitions, and we collect \emph{all} available versions for each submission.
Through this process, we mine 51,743 \nbs from 75 \Kaggle competitions.
Our data cutoff is May 31, 2025; the oldest \nb in our corpus dates back to September 2016.
}

\begin{wrapfigure}{r}{0.45\linewidth}
    \centering
    \includegraphics[width=0.95\linewidth]{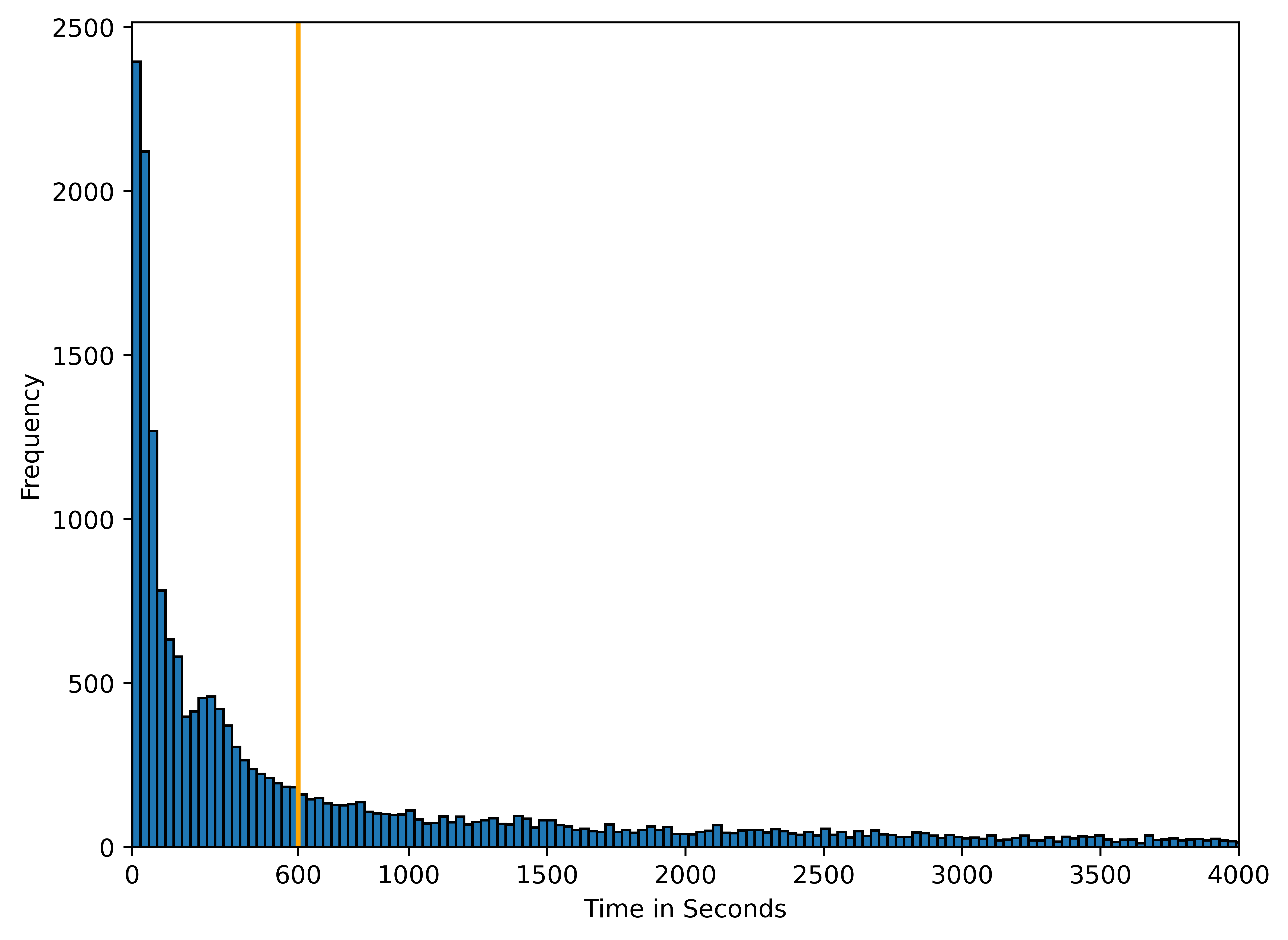}
    \vspace{-10pt}
    \caption{Distribution of \nb runtime.}
    \label{fig:distribution_of_script_time}
    \vspace{-10pt}
\end{wrapfigure}

\paragraph{Filtering Criteria}
We apply several filtering criteria to focus our study on \nbs where \reproducibility can be meaningfully measured. 
First, we retain only \nbs that have been successfully executed on \Kaggle and have a non-zero \score, as these represent \nbs that successfully produced valid predictions and can be evaluated against ground truth.
Second, we retain \nbs with a runtime of less than or equal to 600 seconds (decided by analyzing the distribution of runtime as shown in Figure~\ref{fig:distribution_of_script_time}),
so that our study covers the majority of \nbs and remains tractable in terms of computational resource usage.
Third, we restrict our analysis to 
\par
\Needspace{0.2\textheight}
\noindent
Python \nbs (excluding R \nbs) to maintain consistency in dependency management and execution.
Finally, we exclude \nbs that use external datasets beyond the competition's provided data. 
Such a filter eliminates external effects on \reproducibility (e.g., datasets that may be inaccessible, or have changed over time) and allows us to focus on code modernization challenges.
%
After applying these filters, we retain a total of \NumNotebookAll \nbs for our study.




\subsubsection{Execution Environment Setup}
\label{sec:motivation:baseline:env}

\Kaggle \nbs are executed inside \dockercontainers~\cite{KaggleRunSetup}, and each execution produces a versioned snapshot that is tied to a particular \container version. 
To align with \Kaggle's environment model and minimize environment-induced discrepancies, we execute each \nb inside a GPU-enabled \container based on \Kaggle's official Python image\footnote{\url{https://github.com/Kaggle/docker-python}}, mounting the competition training and test sets provided by \MLEBench.
The same execution environment is used for all experiments in this paper.

\paragraph{Environment and Hyper-Parameters} We conduct our experiments on a server with AMD EPYC 7343 CPU (16 cores @ 1.5GHz), NVIDIA RTX A6000 GPU (48G of RAM built with cuda\_11.5.r11.5), running Ubuntu 22.04 LTS. 
To align with \Kaggle's technical specifications~\cite{KaggleTechSpecifications}, each run is provisioned with 4 CPU cores, 30\,GB of memory, and one GPU.


\paragraph{Containerized Execution and Runtime Measurement}
\Nbs are executed non-interactively via a \nb execution engine inside the \container, with error-tolerant execution enabled so that exceptions do not immediately abort the run. 
Allowing errors ensures we can capture partial progress, logs, and all error locations; some errors (e.g., ones that arise during data visualization) may not affect the final execution outcome and may even be intentionally left unfixed.
We impose a wall-clock cutoff of 600 seconds per \nb, consistent with our data filtering criteria in \S\ref{sec:motivation:baseline:data}.
Runtime is measured only during \nb execution, excluding the overhead of \container initialization, filesystem mounting, and results grading.

\paragraph{Execution Outcome and Grading}
\Kaggle \nbs are expected to produce a competition ``submission'' file in \csv format.
The required schema (\eg column names) is specified by each competition. 
We treat successful \csv generation as the primary execution artifact and evaluate generated submissions using a grader adapted from \MLEBench.
Because \Kaggle's full evaluation benchmark is not publicly released for many competitions, the \MLEBench grader approximates \Kaggle evaluation by splitting the original \Kaggle training set into training and held-out test sets.
For each \nb, we report three metrics: 
(1)~output status (whether the \nb has produced a valid \csv artifact), 
(2)~\reproducedscore, and 
(3)~runtime.
Together, these metrics capture whether the \nb runs end-to-end and whether it produces a meaningful and competitive prediction file.

Reproduction aims to replicate the historical performance that each notebook originally achieved.
Because \MLEBench uses an offline grader (with train/test splits that may differ from \Kaggle's private leaderboard evaluation), \reproducedscores may not exactly match the \targetscores reported on \Kaggle even when the pipeline is correct.
\revise{
We apply a two-sided one-sample $t$-test to assess whether repeated \reproducedscores provide evidence of a mean difference from the \targetscore, using up to ten executions per \nb.
We operationally classify a \nb as \reproducible when the test does not detect such a difference at $\alpha = 0.05$ ($p \geq 0.05$).
When the \reproducedscores are deterministic, making the $t$-test uninformative, we instead apply a heuristic threshold of \aThreshold = \PerfThreshold to classify the \nb as \reproducible if the \reproducedscore differs from the \targetscore by no more than this threshold.
}


\paragraph{Results}

\begin{wraptable}[12]{r}{0.55\linewidth}
\vspace{-8pt}
\begin{small}
\begin{center}
\caption{\Nb \reproducibility results in the baseline environment.}
\label{tab:baseline}
\vspace{-6pt}
\begin{tabular}{|l|l|r|}
\hline
Error Status & Reproducibility & Count \\ \hline
\rowcolor[HTML]{E0E0E0}
\cellcolor[HTML]{E0E0E0}                             & \rep                  & \revise{1,271} \\ \cline{2-3}
\cellcolor[HTML]{E0E0E0}                             & \cellcolor[HTML]{E0E0E0}\nrep{} (w/ CSV)  & \revise{580}   \\ \cline{2-3}
\multirow{-3}{*}{\cellcolor[HTML]{E0E0E0}Error-free} & \cellcolor[HTML]{E0E0E0}\nrep{} (w/o CSV) & \revise{56}    \\ \hline
\rowcolor[HTML]{E0E0E0}
\cellcolor[HTML]{E0E0E0}                             & \rep                  & \revise{1,900} \\ \cline{2-3}
\cellcolor[HTML]{E0E0E0}                             & \cellcolor[HTML]{E0E0E0}\nrep{} (w/ CSV)  & \revise{1,466} \\ \cline{2-3}
\multirow{-3}{*}{\cellcolor[HTML]{E0E0E0}Error}      & \cellcolor[HTML]{E0E0E0}\nrep{} (w/o CSV) & \revise{6,536} \\ \hline
\multicolumn{2}{|l|}{Failed (timeout)}                           & \revise{275}   \\ \hline
\multicolumn{2}{|l|}{Failed (notebook not saved)}                & \revise{22}    \\ \hline
\end{tabular}

\end{center}
\end{small}
\vspace{-10pt}
\end{wraptable}

Table~\ref{tab:baseline} shows baseline outcomes by execution status and \reproducibility, with output status (whether the \nb has produced a valid \csv artifact) indicated for \nonreproducible \nbs.
Among the \revise{12,106} \nbs in the study, \revise{3,171 (26\%)} are classified as \reproducible.
Among error-free \nbs, \revise{1,271} are \xEFR, \revise{636} are \xEFNR, of which \revise{56} do not generate a \csv result.
The latter can occur because some submissions produce artifacts that do not match the grader's expected \csv format or naming conventions, or because the collected version does not contain the full submission-generation code.
Among \nbs that raise errors, \revise{1,900} are \xER, \revise{8,002} are \xENR (where \revise{6,536} have no \csv).
An additional \revise{297} \nbs are classified as failed (all ten execution attempts fail): \revise{275} time out before completion and \revise{22} correspond to notebook versions whose outputs were not written back to the \nb.

\conclusionbox{
\noindent\textbf{\underline{Summary}:} 
Under the contemporary baseline environment, only \revise{26\%} of the studied Kaggle \nbs reproduce; the majority are non-reproducible due to execution errors, missing \csv output, or score deviation, and a small fraction fail due to timeouts or missing saved outputs.
}

\subsection{Environment Backporting}
\label{sec:motivation:backporting}
Backporting aims to reconstruct the historical software environment that originally executed each \nb on \Kaggle.
Since \Kaggle's \dockerimages older than two years are no longer available for download, we approximate historical environments by downgrading dependencies to versions consistent with each notebook's submission timestamp.

\subsubsection{Backporting Algorithm}
\label{sec:motivation:backporting:env}

We design a rule-based procedure for selecting historical versions of Python and third-party libraries based on submission metadata and code characteristics.

\paragraph{Dependency Analysis}
To identify the dependencies used by each notebook, we apply an AST-based dependency extraction tool (pigar~\cite{pigar}) to the notebook code and generate dependency lists with correct PyPI package names.
These dependency lists, together with the submission timestamp, form the inputs to our rule-based backporting procedure.

\paragraph{Inferring Python Version} 
%
First, we infer the Python major version (2 vs.\ 3) and minor version by cross-referencing the notebook's submission timestamp with the release dates of historical Python versions.
The code is also analyzed to detect syntax patterns characteristic of Python~2 (for example, bare \texttt{print} statements, \texttt{xrange}, \texttt{raw\_input}) or Python~3 (for example, \texttt{print()} calls, f-strings, \texttt{async}/\texttt{await}). 
We choose the latest \CodeIn{major.minor} version series that predates the submission but always use the latest patch release available within that series, because it minimizes dependency conflicts, and newer patch versions surpass older ones without breaking changes.
For example, for a notebook submitted on 2021-10-01, we will first select Python 3.9 released on 2020-10-05 instead of Python 3.10 released on 2021-10-04, then select Python 3.9.25 that is released on 2025-10-31.

\paragraph{Selecting Dependency Versions}
Given a \nb and its inferred Python version, we select historical versions of all third-party dependencies used by that \nb. 
For each imported package, we query the corresponding package index to retrieve the full release history, along with per-release metadata, required Python versions, and yanked-status flags. 
We filter out yanked releases and, for each remaining version, record both the release time and the declared Python compatibility.
For a given \nb, we then choose the newest version released before the submission timestamp; if no such version exists, we fall back first to the oldest version that is compatible with the inferred Python version and, as a last resort, to the oldest available version.
The resulting per-notebook dependency lists are written in a requirements-style format that constrains each package to be less than or equal to the selected version (\eg \texttt{package\_name<=v\_hist}). 
Our rule-based selection strategy is designed to be robust, as it systematically respects submission dates, enforces Python-version compatibility, 
and promotes the use of shared minimal versions.

\paragraph{Environment Materialization}
%
We materialize the inferred environments based on the baseline \dockercontainer (\S\ref{sec:motivation:baseline:env}), using \CodeIn{virtualenv} to install the inferred Python version and the selected dependency versions.
If the backported environment cannot be installed (e.g., due to dependency conflicts or unavailable historical wheels), we mark the \nb as Failed.

\subsubsection{Reproducibility Analysis}
\label{sec:motivation:backporting:result}

\begin{wraptable}[13]{r}{0.55\linewidth}
\vspace{-8pt}
\begin{small}
\begin{center}
\caption{\Nb \reproducibility results in the backported environment.}
\label{tab:downgrade}
\vspace{-6pt}
\begin{tabular}{|l|l|r|}
\hline
Error Status & Reproducibility & Count \\ \hline
\rowcolor[HTML]{E0E0E0}
\cellcolor[HTML]{E0E0E0}                             & \rep                  & \revise{972}   \\ \cline{2-3}
\cellcolor[HTML]{E0E0E0}                             & \nrep{} (w/ CSV)       & \revise{817}   \\ \cline{2-3}
\multirow{-3}{*}{\cellcolor[HTML]{E0E0E0}Error-free} & \nrep{} (w/o CSV)      & \revise{62}    \\ \hline
\rowcolor[HTML]{E0E0E0}
\cellcolor[HTML]{E0E0E0}                             & \rep                  & \revise{515}   \\ \cline{2-3}
\cellcolor[HTML]{E0E0E0}                             & \nrep{} (w/ CSV)       & \revise{1,900} \\ \cline{2-3}
\multirow{-3}{*}{\cellcolor[HTML]{E0E0E0}Error}      & \nrep{} (w/o CSV)      & \revise{7,251} \\ \hline
\multicolumn{2}{|l|}{Failed (timeout)}                           & \revise{215}   \\ \hline
\multicolumn{2}{|l|}{Failed (notebook not saved)}                & \revise{9}     \\ \hline
\multicolumn{2}{|l|}{Failed (backporting)}                       & \revise{365}   \\ \hline
\end{tabular}
\end{center}
\end{small}
\vspace{-10pt}
\end{wraptable}

\paragraph{Results}

Table~\ref{tab:downgrade} shows \backporting outcomes by execution status and \reproducibility.
Among error-free \nbs, \revise{only 972} are \xEFR, and \revise{879} are \xEFNR, of which \revise{62} do not generate a \csv.
Among \nbs that raise errors, \revise{merely 515} are \reproducible (\xER); \revise{9,151} are \xENR (where \revise{7,251} have no \csv).
An additional \revise{589} \nbs are classified as failed due to timeouts, unsaved \nb outputs, or failures to materialize a viable \backporting environment (\eg incompatible transitive dependencies, unavailable historical wheels, or exceeding the 30-minute setup limit).

\begin{figure}[t]
    \centering
    \includegraphics[width=.8\linewidth]{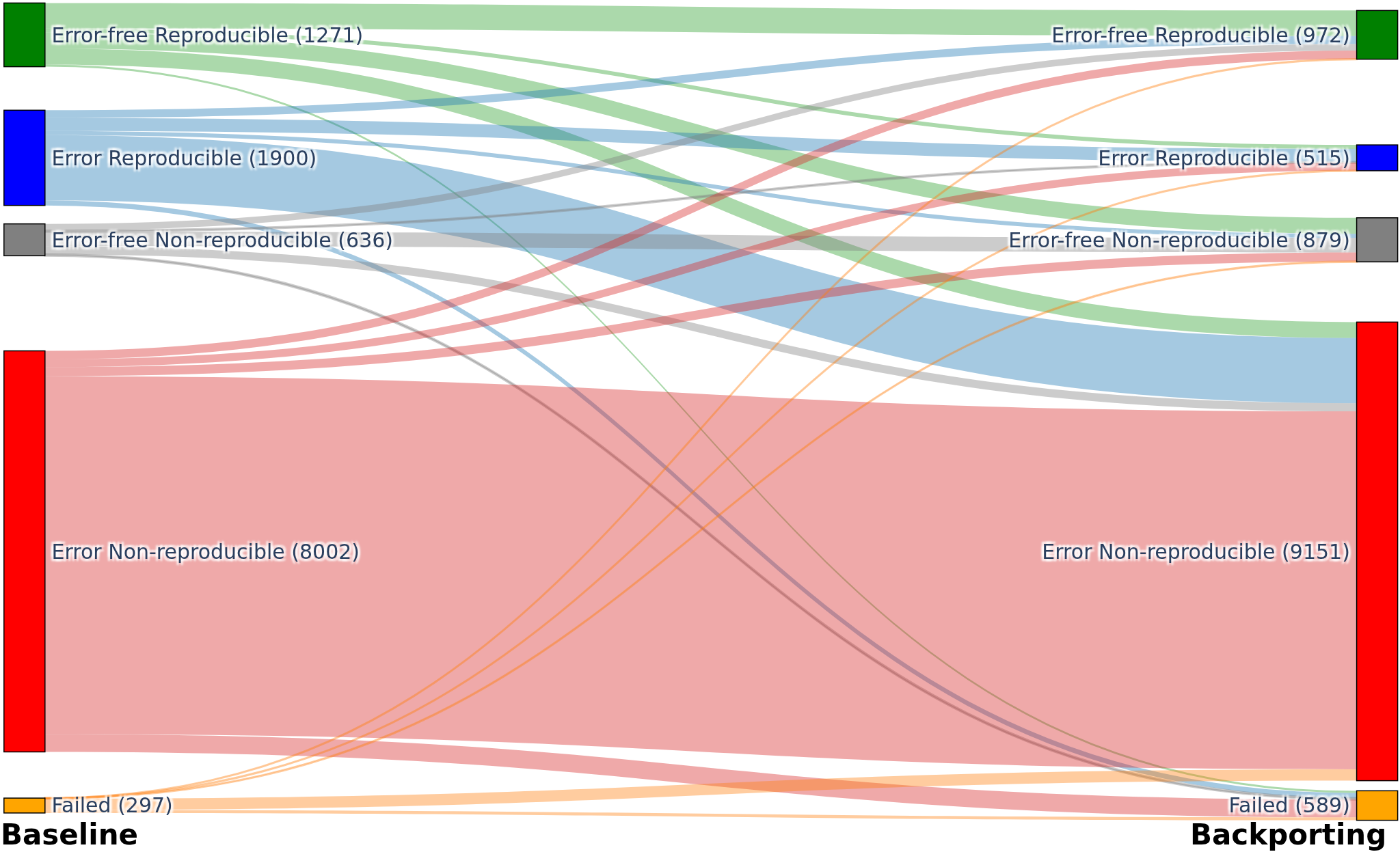}
    \vspace{-10pt}
    \caption{\revise{\Nb reproducibility flow from \bsl to \dg.}}
    \label{fig:sankey_baseline_backporting}
\end{figure}

\paragraph{Comparison with Baseline}
Figure~\ref{fig:sankey_baseline_backporting} illustrates the flow of \nbs from \bsl to \dg: at \bsl, \revise{1,271} are \xEFR, \revise{1,900} are \xER, \revise{636} are \xEFNR, \revise{8,002} are \xENR, and \revise{297} are failed.
Backporting reduces the number of \xEFR \nbs (\revise{1,271} to \revise{972}) and nearly doubles the failed set (\revise{297} to \revise{589}), while \xENR jumps (\revise{8,002} to \revise{9,151}); many \nbs that were \reproducible at \bsl become \nonreproducible or fail under the backported environment, and a substantial flow from \xENR enters the failed category.

\textbf{Backporting, therefore, does not improve \reproducibility and introduces more failure modes.} 
The results imply that simply downgrading dependencies to match submission timestamps is not a reliable remedy when migrating \nbs or reusing scripts from other sources in one's own environment: many \nbs remain \nonreproducible or become outright failures due to environment reconstruction issues, so backporting is not recommended as a general-purpose fix.

\conclusionbox{
\noindent\textbf{\underline{Summary}:}
Under the backported environment, \revise{1,487} \nbs (\revise{12\%}) are \reproducible, fewer than the baseline (\revise{3,171, i.e., 26\%}). 
\Envbackporting does not improve \reproducibility and introduces additional failure modes; downgrading dependencies to match submission timestamps is not a reliable fix when migrating or reusing MLE \nbs.
}



\section{\Tool Technique}
\label{sec:technique}

\begin{figure}[t]
\centering
\includegraphics[width=.95\textwidth]{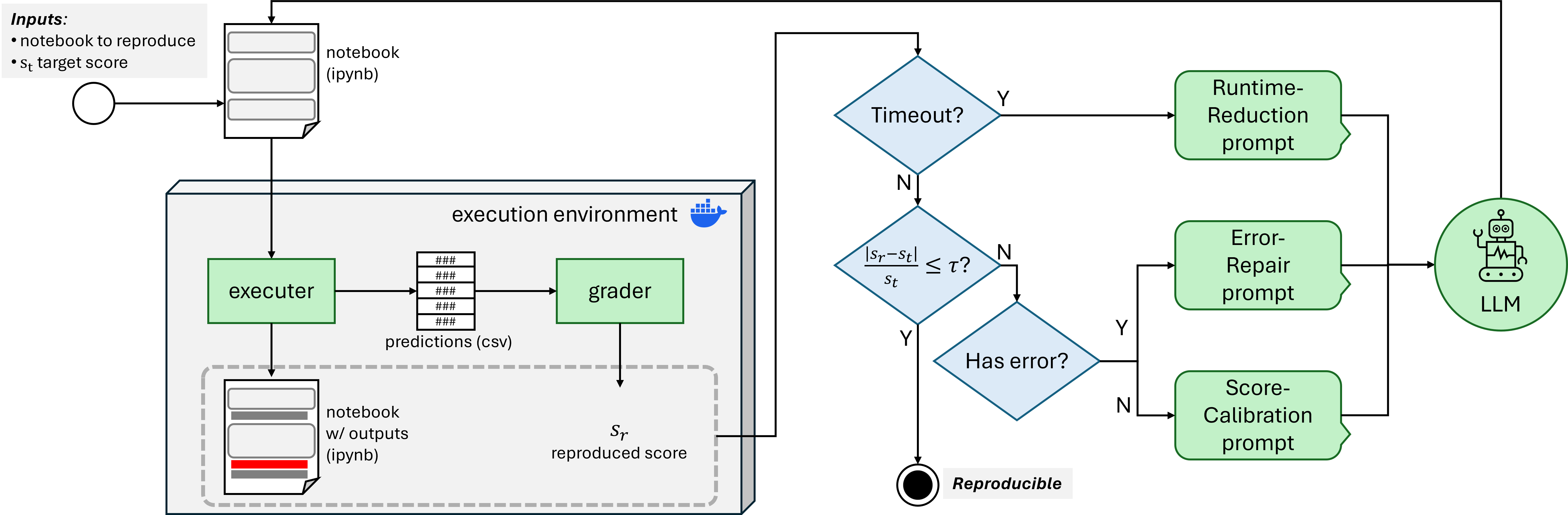}
\vspace{-6pt}
\caption{\Tool's workflow.}
\label{fig:workflow}
\end{figure}

Because \envbackporting fails to address the \reproducibility gap, and MLE execution environments evolve rapidly, we \emph{modernize} \nb code while keeping the shared base container and hardware configuration fixed.
We design \Tool as an LLM-driven agentic framework that iteratively fixes a \nb toward reproducibility.
LLMs have shown promising capabilities for program repair and code migration, but unlike prior work, the LLM fixes in \Tool need to be performance-aware: \Tool targets not only successful execution, but also reproducing the originally reported \score.
Thus, \Tool grounds LLM fixes with rich execution feedback (runtime, error tracebacks, and score deviation) and uses specific prompts (\runtimereduction, \errorrepair, \scorecalibration) to guide each LLM fix.
An overview of \Tool's workflow is shown in Figure~\ref{fig:workflow}.



\subsection{Agentic Workflow}
\label{sec:technique:workflow}

\Tool takes as inputs a \nb and a \targetscore \aTargetScore to be reproduced.
Then, \Tool adopts an agentic workflow that iteratively collects execution feedback and applies LLM-driven fixes.

At each iteration, \Tool performs the following steps:
\begin{enumerate}[topsep=3pt,itemsep=1ex,partopsep=0ex,parsep=0ex,leftmargin=*]
\item executes the \nb and collects execution feedback, namely the outputs, error tracebacks (if any error occurs), \reproducedscore{} \aReproducedScore, and runtime (detailed in \S\ref{sec:technique:execution});
\item analyzes the execution feedback to detect one of three \nonreproducibility issues:
\begin{itemize}[noitemsep,topsep=2pt]
\item if the execution times out, which can be caused by API misuse or not enabling GPU acceleration, \Tool first needs to reduce the runtime;
\item if the reproduced score deviates from the target score, and the \nb has errors, \Tool attempts to fix the errors as they are likely to be the root causes of \nonreproducibility;
\item if the score deviates, and the \nb does not have errors, a hidden API behavior change may be responsible, and \Tool needs to calibrate the score to the target.
\end{itemize}
\item triggers a targeted LLM fix based on the detected issue (detailed in \S\ref{sec:technique:llm}).
\end{enumerate}

\Tool terminates when the \nb becomes \reproducible (i.e., the score deviation is within the \revise{heuristically chosen} replication threshold \aThreshold = \PerfThreshold).
We also set a maximum number of iterations \aMaxIterations = 16; if the \nb does not become \reproducible within the cap, it is assigned a terminal outcome (e.g., \xENR, \xEFNR, or Failed due to timeout/LLM failure).

\subsection{Execution Feedback Collection}
\label{sec:technique:execution}

Each notebook is executed inside the containerized environment as described in \S\ref{sec:motivation:baseline:env}.
We also mount the \MLEBench competition dataset into the container so \nbs can access inputs and emit the submission \csv to standardized paths.

The base container is shared across all \nbs, while notebook-authored dependency setup remains part of the \nb code.
The executor therefore applies any \CodeIn{!pip install} commands before executing the remaining cells; these commands can change packages for that \nb's run without changing the shared starting container.
Then, the \nb is executed in an error-tolerant mode that continues past failing cells; the outputs and all error tracebacks are collected and saved to the \nb file.
If the ML pipeline runs successfully, a submission \csv should be produced.
The executor has a wall-clock time limit of 600 seconds.
The grader, which is adapted from \MLEBench~\cite{Chan24MLEBench}, grades the submission \csv to obtain the \reproducedscore \aReproducedScore.

\subsection{LLM Fixing}
\label{sec:technique:llm}

The input to the LLM is a structured prompt that includes a context-specific part and a shared part.
The context-specific part depends on the detected \nonreproducibility issue and instructs the LLM to follow certain requirements when performing the fix.
There are three types of fixes: \RuntimeReduction when the execution times out, \ErrorRepair when the \nb has errors, and \ScoreCalibration when the \nb does not have errors.

The shared part contains the following information:
\begin{itemize}[topsep=3pt,itemsep=1ex,partopsep=0ex,parsep=0ex,leftmargin=*]
  \item \textbf{Kaggle task description}: describing the competition's task, evaluation metric, and dataset schema.
  \item \textbf{Execution environment}: describing the Python version and dependency libraries' versions.
  \item \textbf{I/O contract}: specifying the paths for input data (\CodeIn{/kaggle/input/}) and the expected submission \csv location (\CodeIn{/kaggle/working/}).
  \item \textbf{Scores}: target \aTargetScore, current \aReproducedScore, and the metric directionality (higher/lower is better).
  \item \textbf{Notebook}: the current notebook rendered in a cell-delimited format; when there are errors, the traceback is appended to the corresponding cell.
\end{itemize}

The LLM is also instructed to return a response in a fixed plan \(\rightarrow\) patch format: 
(1)~a short plan that enumerates the intended changes and their expected effect on the observed failure mode, followed by 
(2)~a single markdown code block containing the full updated notebook in the same cell-delimited representation.
\Tool parses this output deterministically and applies the patch to the \nb for the next iteration.
This protocol both enforces a concrete, executable patch and preserves sufficient structure for automated application and auditing.

\subsection{Running Example}
\label{sec:technique:example}

\begin{figure}[t]
\centering
\includegraphics[width=.96\textwidth]{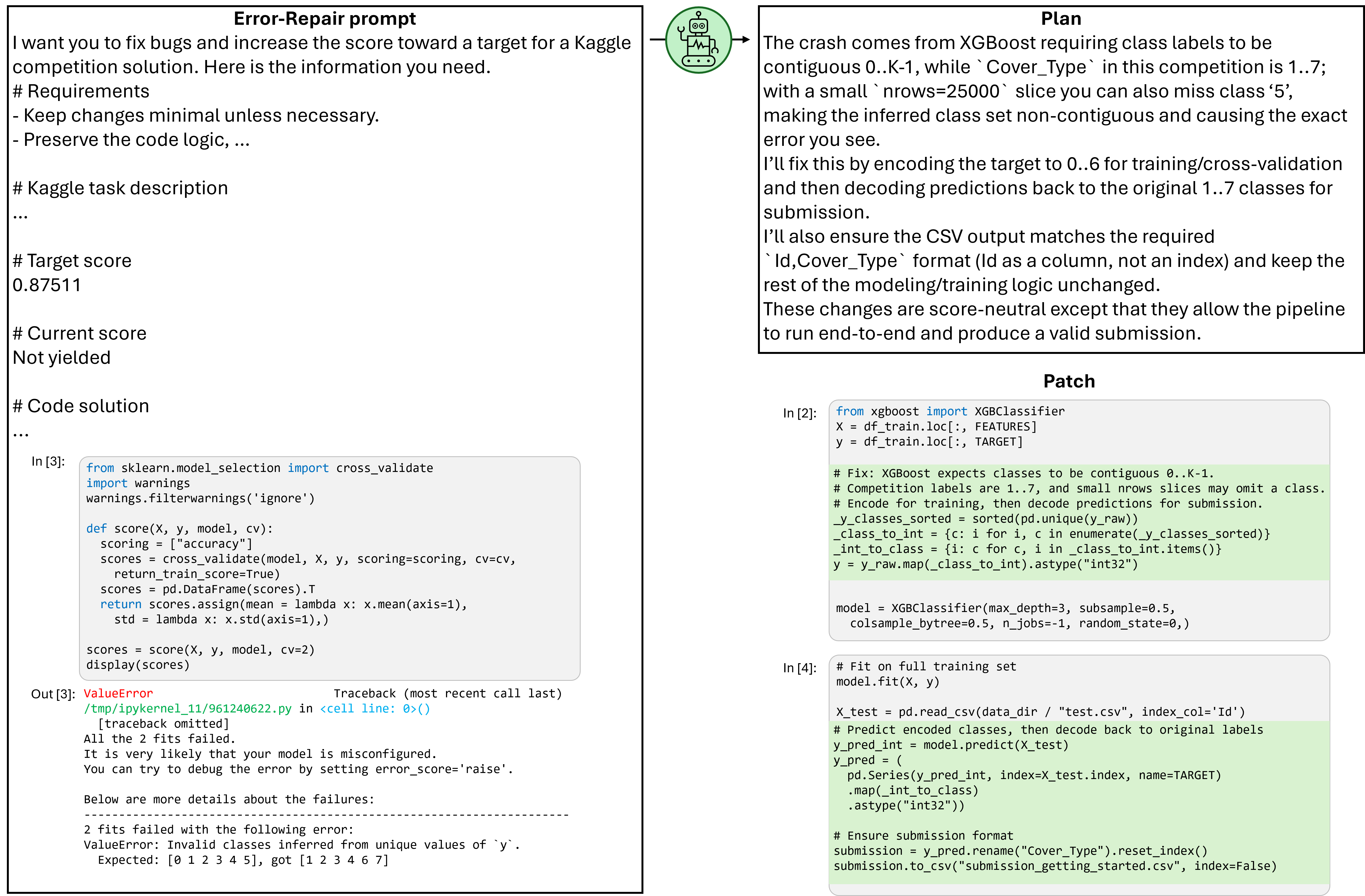}
\vspace{-10pt}
\caption{\Tool successfully modernizes the \nb in Figure~\ref{fig:motivation-example} (\aReproducedScore = 0.93778 compared to \aTargetScore = 0.87511), by re-encoding the dataset's classes (1..7) as the labels required by XGBoost (0..6).}
\label{fig:example-prompt-patch}
\end{figure}

In the motivating example in Figure~\ref{fig:motivation-example}, the execution fails under a newer XGBoost version because the classifier expects labels in \(\{0,\dots,K-1\}\) but the dataset provides labels in \(\{1,\dots,7\}\).
\Tool detects this failure from the traceback, issues an \ErrorRepair prompt, and applies a patch that inserts an explicit label re-encoding step (e.g., shifting/encoding labels to start at 0) while preserving the original training/evaluation logic and submission format.
As shown in Figure~\ref{fig:example-prompt-patch}, the patched \nb executes successfully and achieves a score of 0.93778, slightly above the \targetscore of 0.87511.

\section{Evaluation}
\label{sec:eval}

\paragraph{Subjects}
\Tool focuses on \nonreproducible \nbs: these notebooks either fail to produce a valid submission artifact or \revise{fail the statistical \reproducibility criterion (one-sample $t$-test, \S\ref{sec:motivation:baseline:env})}.
We exclude \reproducible \nbs because they already run and reproduce under the baseline environment and offer little headroom for improvement.
We also exclude \nbs that hard-fail due to timeouts or pathological resource usage, as such cases are dominated by system-level constraints rather than code-level modernization.

From the \revise{8,638} \nonreproducible \nbs identified in the baseline (\S\ref{sec:motivation:baseline}), we further filter candidates based on prompt length.
For each notebook, \Tool constructs prompts that include the source code (organized into cells) and, when applicable, error tracebacks from previous executions. To avoid exceeding the context window of the LLM and to prevent repeated failures due to overlong prompts, we empirically select a token cutoff based on the joint distribution of tokenized source and error text. 
Specifically, we tokenize all \nbs and associated error messages using an OpenAI model-compatible tokenizer~\cite{tiktoken}, compute the 95th percentile of the resulting token counts, and obtain a cutoff of \revise{16,337} tokens. 
Notebooks whose combined source and error text exceed this cutoff are excluded from the pool.
The final subject set consists of \revise{\NumNotebookSubject} \nbs (out of \revise{8,638} \nonreproducible \nbs); we remove \revise{428} extremely large notebooks to keep LLM calls within context limits and avoid disproportionate allocation of the compute budget.

\paragraph{Experimental Setup} We use the same software and hardware setup as the motivation study (\S\ref{sec:motivation:baseline:env}).
We use OpenAI's GPT-5.2 (the 2025-12-11 version) as the LLM in \Tool \revise{and the open-weight GPT-OSS-120b for comparison}.

\paragraph{Manual Inspection} \revise{We statistically sample a subset (368 \{before, after\} \nb pairs modernized by \Tool, ensuring a 95\% confidence level and 5\% margin of error) and manually verify crash types and crash phases within the ML pipeline, as outlined in \citet{WangETAL25NotebookCrash}'s work. In addition, functional equivalence of the \nb pairs before and after upgrade is also inspected. Specifically, two annotators (the first author and a Master's student in CS) independently label the \nb pairs, following a closed coding procedure~\cite{SeamanTSE99ClosedCodingProcedure}. Then, the two annotators discuss the labels for each \nb pair on which they disagree and establish consensus. If any conflicts persist, a third annotator (who is a computer science Professor in MLE and also an author of this paper) arbitrates remaining conflicts. Inter-rater reliability (IRR) is measured by Cohen's kappa~\cite{Cohen60kappa}.}

\paragraph{Research Questions}
We study the following research questions:

\DefMacro{rq-reproducibility}{RQ1}
\noindent
\MyPara{\UseMacro{rq-reproducibility}} How many \nbs become \reproducible after applying \Tool?

\DefMacro{rq-fix-num-type}{RQ2}
\noindent
\MyPara{\UseMacro{rq-fix-num-type}} How many and what kinds of fixes does the LLM perform?

\DefMacro{rq-similarity}{RQ3}
\noindent
\MyPara{\UseMacro{rq-similarity}} How much code modification is needed to make the \nbs reproducible?

\DefMacro{rq-cost}{RQ4}
\noindent
\MyPara{\UseMacro{rq-cost}} How much does it cost to apply \Tool?

\DefMacro{rq-error-type}{RQ5}
\noindent
\MyPara{\UseMacro{rq-error-type}} Which types of errors can and cannot be fixed by \Tool?


\subsection{\UseMacro{rq-reproducibility}: \Tool Reproducibility}
\label{sec:eval:rq-reproducibility}
Table~\ref{tab:file-upgrade} reports file-level modernization outcomes \revise{for the proprietary GPT-5.2 and open-weight GPT-OSS-120b models}, categorized by execution status and \reproducibility.
With \Tool \revise{supported by GPT-5.2}, \FileReproducibleNumNotebook of the \NumNotebookSubject upgraded \nbs (\FileReproducibility) become \reproducible: \revise{2,397} end-to-end runs complete without errors and pass the one-sample $t$-test (\xEFR), and \FileErrorReproducibleNumNotebook produce a \reproducible outcome despite residual execution errors (\xER).

\begin{table}[t]
\begin{small}
\centering
\caption{Reproducibility after applying \Tool to the \NumNotebookSubject \nonreproducible \nbs.}
\label{tab:file-upgrade}
\vspace{-8pt}
\begin{subtable}[t]{0.48\textwidth}
\centering
\caption{Upgrade with GPT-5.2.}
\label{tab:gpt-file-upgrade}
\vspace{-2pt}
\resizebox{\linewidth}{!}{
\begin{tabular}{|l|l|r|}
\hline
Error Status & Reproducibility & Count \\ \hline
\rowcolor[HTML]{E0E0E0}
\cellcolor[HTML]{E0E0E0}                             & \rep                  & \revise{2,397} \\ \cline{2-3}
\cellcolor[HTML]{E0E0E0}                             & \nrep{} (w/ CSV)       & \revise{2,578} \\ \cline{2-3}
\multirow{-3}{*}{\cellcolor[HTML]{E0E0E0}Error-free} & \nrep{} (w/o CSV)      & \revise{1}     \\ \hline
\rowcolor[HTML]{E0E0E0}
\cellcolor[HTML]{E0E0E0}                             & \rep                  & \revise{895}   \\ \cline{2-3}
\cellcolor[HTML]{E0E0E0}                             & \nrep{} (w/ CSV)       & \revise{633}   \\ \cline{2-3}
\multirow{-3}{*}{\cellcolor[HTML]{E0E0E0}Error}      & \nrep{} (w/o CSV)      & \revise{243}   \\ \hline
\multicolumn{2}{|l|}{Failed (timeout)}                           & \revise{1,023} \\ \hline
\multicolumn{2}{|l|}{Failed (notebook not saved)}                & \revise{424}   \\ \hline
\multicolumn{2}{|l|}{Failed (LLM failure)}                       & \revise{16}    \\ \hline
\end{tabular}
}
\end{subtable}
\hfill
\begin{subtable}[t]{0.48\textwidth}
\centering
\caption{\revise{Upgrade with GPT-OSS-120b.}}
\label{tab:oss-file-upgrade}
\vspace{-2pt}
\revise{\resizebox{\linewidth}{!}{
\begin{tabular}{|l|l|r|}
\hline
Error Status & Reproducibility & Count \\ \hline
\rowcolor[HTML]{E0E0E0}
\cellcolor[HTML]{E0E0E0}                             & \rep                  & 2,877 \\ \cline{2-3}
\cellcolor[HTML]{E0E0E0}                             & \nrep{} (w/ CSV)       & 2,348 \\ \cline{2-3}
\multirow{-3}{*}{\cellcolor[HTML]{E0E0E0}Error-free} & \nrep{} (w/o CSV)      & 22    \\ \hline
\rowcolor[HTML]{E0E0E0}
\cellcolor[HTML]{E0E0E0}                             & \rep                  & 806   \\ \cline{2-3}
\cellcolor[HTML]{E0E0E0}                             & \nrep{} (w/ CSV)       & 303   \\ \cline{2-3}
\multirow{-3}{*}{\cellcolor[HTML]{E0E0E0}Error}      & \nrep{} (w/o CSV)      & 371   \\ \hline
\multicolumn{2}{|l|}{Failed (timeout)}                           & 1,060 \\ \hline
\multicolumn{2}{|l|}{Failed (notebook not saved)}                & 416   \\ \hline
\multicolumn{2}{|l|}{Failed (LLM failure)}                       & 7     \\ \hline
\end{tabular}
}}
\end{subtable}
\end{small}
\end{table}

\begin{figure}
    \centering
    \includegraphics[width=0.8\linewidth]{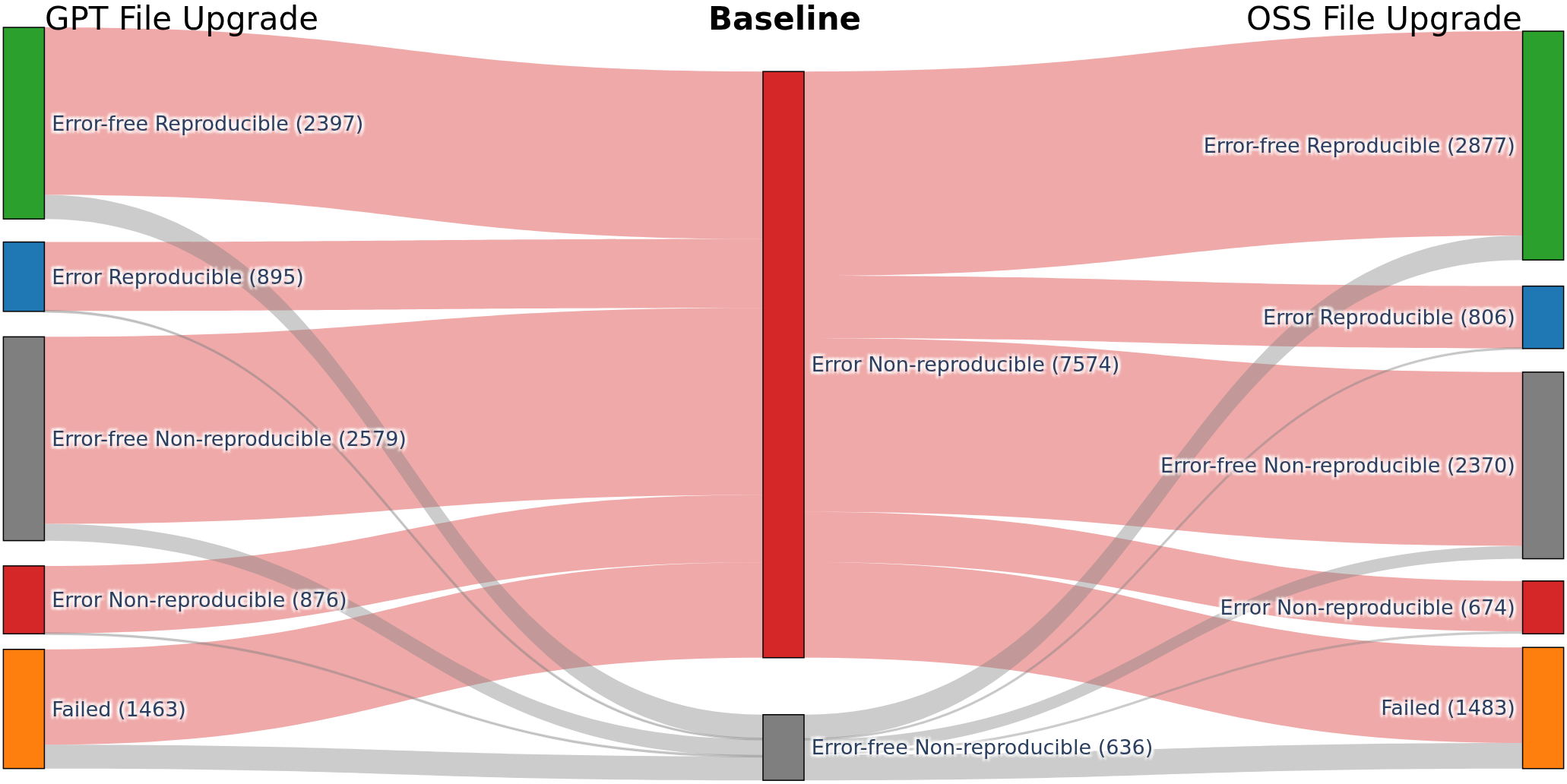}
    \vspace{-6pt}
    \caption{\revise{\Nb \reproducibility transitions from \bsl to \Tool modernized \nb.}}
    \label{fig:sankey_baseline_file}
\end{figure}


Among \revise{2,579} \xEFNR \nbs, 2,578 are w/ \csv, and \revise{1} is w/o \csv generated.
Among \nbs that raise errors, \revise{876} are \xENR (where \revise{243} have no \csv).
An additional \revise{1,463} \nbs are classified as failed.
Specifically, Figure~\ref{fig:sankey_baseline_file} (left) illustrates the flow from \bsl to file-level upgrade: at \bsl, \revise{636} \xEFNR and \revise{7,574} \xENR \nbs are fed into the modernization pool (\NumNotebookSubject in total); after upgrading, \revise{2,397} become \xEFR, \revise{895} become \xER, \revise{2,579} remain \xEFNR, \revise{876} remain \xENR, and \revise{1,463} fail.

\revise{
Using the open-weight GPT-OSS-120b, 3,683 of the \NumNotebookSubject upgraded \nbs (44.9\%) become \reproducible (Table~\ref{tab:oss-file-upgrade}, Figure~\ref{fig:sankey_baseline_file} (right)): 2,877 end-to-end runs complete without errors and pass the one-sample $t$-test (\xEFR), and 806 produce a \reproducible outcome despite residual execution errors (\xER).
Among 2,370 \xEFNR \nbs, 2,348 have \csv generated, and 22 are w/o \csv generated; among \nbs that raise errors, 674 are \xENR (371 w/o \csv).
An additional 1,483 \nbs fail (1,060 timeouts, 416 unsaved outputs, and 7 LLM failures).
Figure~\ref{fig:sankey_baseline_file} (right) shows a qualitatively similar transition pattern from \bsl to file-level upgrade: after modernization, 2,877 become \xEFR, 806 become \xER, 2,370 remain \xEFNR, 674 remain \xENR, and 1,483 fail.}

The upgraded cohort is drawn from \nrep \nbs in \bsl (\S\ref{sec:motivation:baseline}); in that cohort, zero \nbs are \reproducible at \bsl.
\Tool raises the share of \reproducible \nbs from 0\% to \FileReproducibility \xspace (\FileReproducibleNumNotebook of \NumNotebookSubject{}) with GPT-5.2, including \revise{2,397} \nbs that are also error-free and \FileErrorReproducibleNumNotebook that remain \reproducible despite residual errors. With GPT-OSS-120b, \Tool achieves a higher \reproducibility rate of 44.9\%, reproducing 3,683 \nbs, including \revise{2,877} \nbs that are error-free and 806 that remain \xER.

\revise{Beyond statistical \reproducibility, we also assess functional equivalence between each baseline \nb and its modernized version (manual inspection protocol above).
Among pairs upgraded by GPT-5.2, 71.5\% are judged functionally equivalent, with an IRR of 0.70 (substantial consensus).
Among the non-equivalent pairs, 47.6\% of mismatches are attributed to \scorecalibration, which is allowed to perform small parameter/model changes to reach the target score under the new environment.
With GPT-OSS-120b, only 52.7\% of pairs are functionally equivalent (IRR = 0.78, substantial consensus); among its non-equivalent pairs, 46.9\% of mismatches are due to \scorecalibration.}

\revise{
GPT-OSS-120b achieves a \reproducibility rate 4.8 percentage points higher than GPT-5.2 (44.9\% vs.\ \FileReproducibility), but upon inspection, GPT-5.2 yields substantially more functionally equivalent modernizations (71.5\% vs.\ 52.7\%).
This shows a dilemma that achieving score-based \reproducibility and preserving semantic parity are not always compatible.
We therefore focus the RQ2 and RQ3 analyses on GPT-5.2 to examine fix statistics and code-edit patterns under the backend that better preserves functional equivalence, which is more important in our goal.
}

\conclusionbox{
\noindent\textbf{\underline{Summary}:}
\revise{\Tool restores statistical \reproducibility for \FileReproducibility of \nbs with GPT-5.2 and 44.9\% with GPT-OSS-120b; the GPT-OSS-120b rate is 4.8 percentage points higher.
Manual inspection finds higher functional consistency with GPT-5.2 (71.5\%, IRR = 0.70) than GPT-OSS-120b (52.7\%, IRR = 0.78); most non-equivalent mismatches stem from \scorecalibration.}
Overall, \Tool restores reproducibility for a large fraction of previously non-reproducible MLE notebooks.
}



\subsection{\UseMacro{rq-fix-num-type}: Statistics of LLM Fixes}
\label{sec:eval:rq-fix-num-type}
\begin{figure}[t]
    \centering
    \begin{subfigure}[t]{0.5\linewidth}
        \centering
        \includegraphics[width=\linewidth]{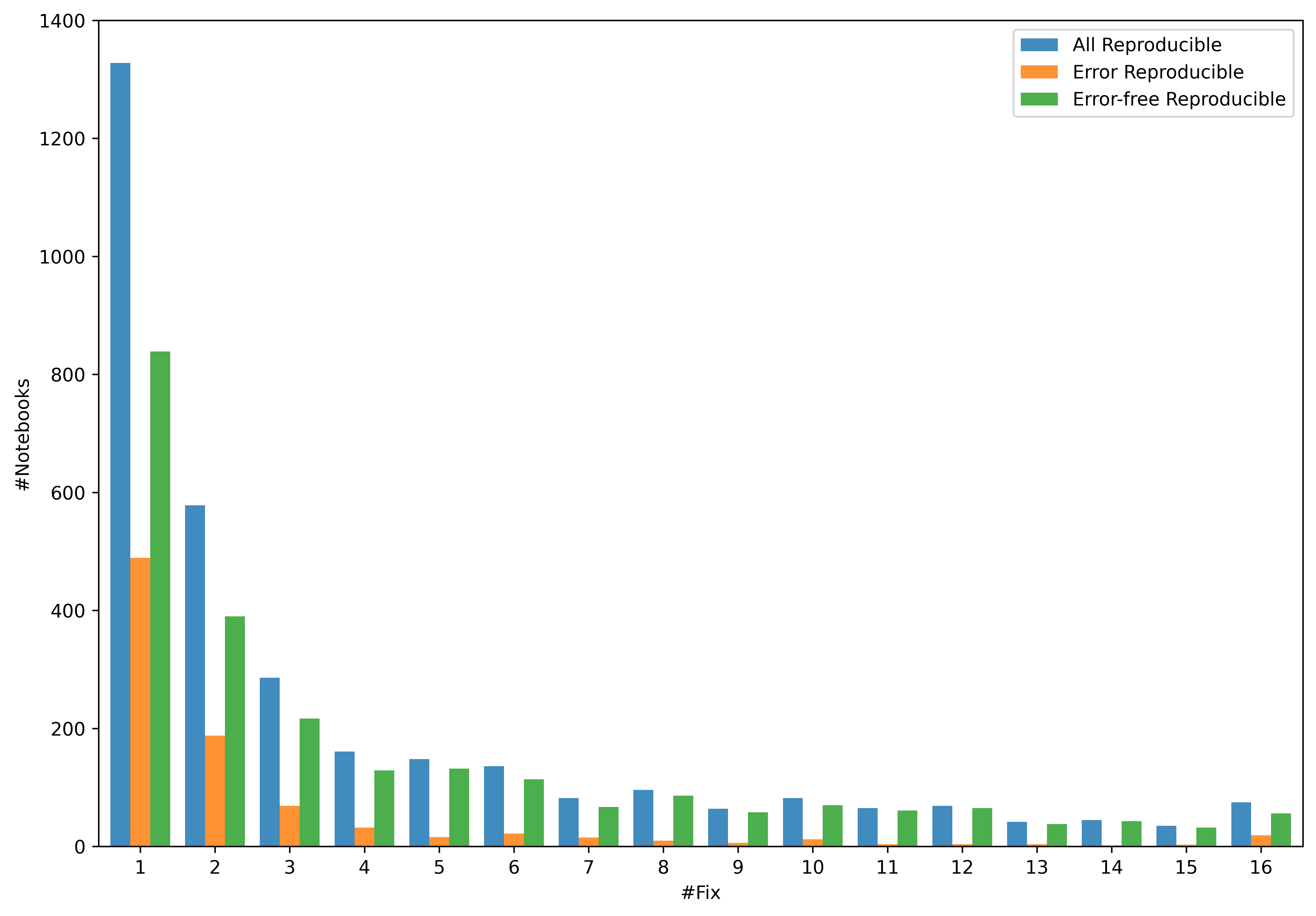}
        \caption{\revise{Number of \nbs that become \reproducible at each fix count.}}
        \label{fig:fix_distribution_file}
    \end{subfigure}%
    \hspace{0.02\linewidth}%
    \begin{subfigure}[t]{0.46\linewidth}
        \centering
        \includegraphics[width=\linewidth]{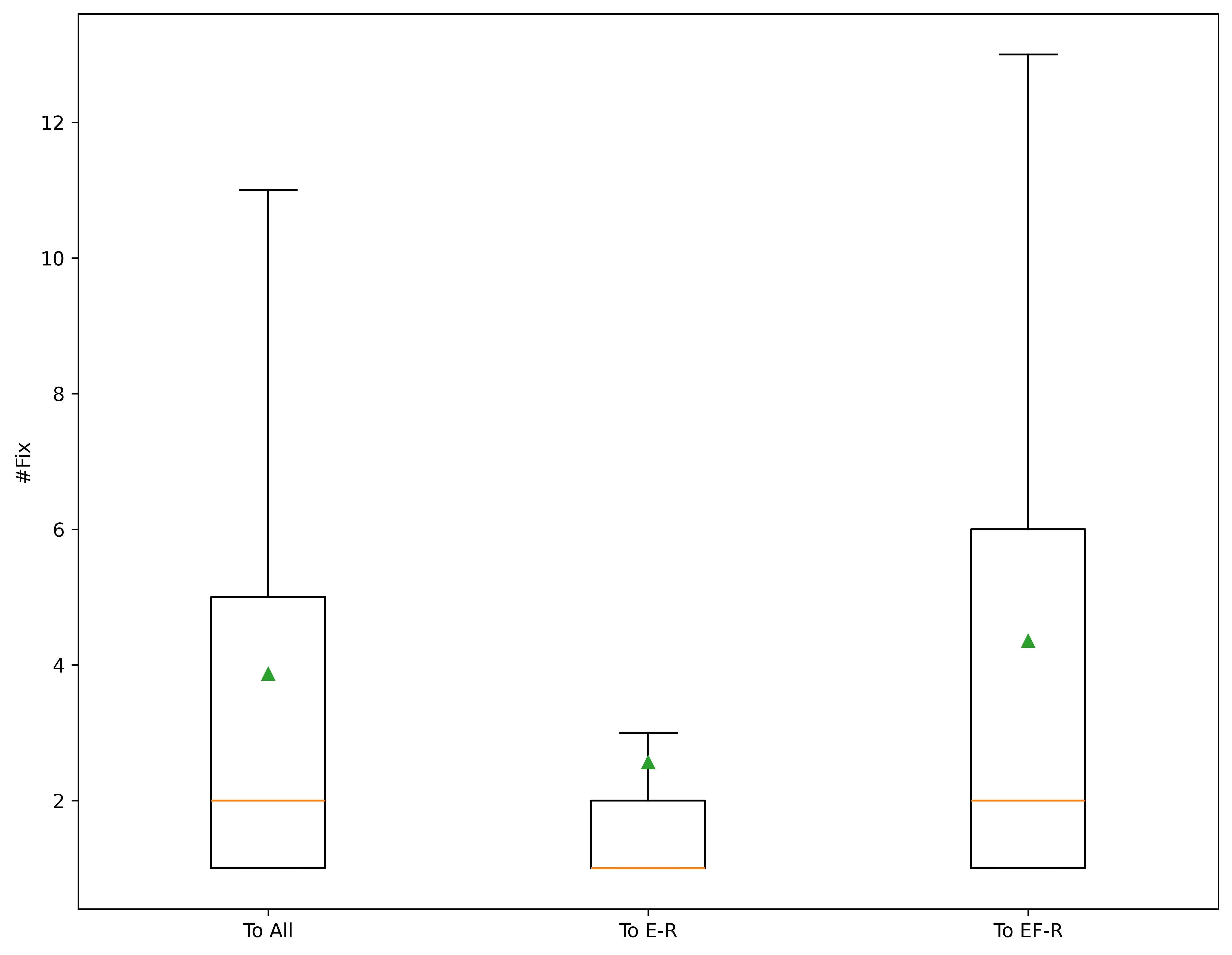}
        \caption{\revise{Box plots of number of LLM fixes.
        ``To All'' represents all \srep \nbs (\UseMacro{TH-error-reproducible} and \UseMacro{TH-error-free-reproducible}) after modernization.}}
        \label{fig:fix_boxplot_file}
    \end{subfigure}
    \vspace{-8pt}
    \caption{\revise{Distribution of the number of LLM (GPT-5.2) fixes performed by \Tool.}}
\end{figure}

Figure~\ref{fig:fix_distribution_file} shows how many \nbs become \reproducible at each fix count. 
The distribution is heavily skewed toward few fixes, with the majority of \nbs becoming \reproducible after 1--3 fixes;
\revise{at 1 fix, which can be seen as a zero-shot LLM repair baseline, the modernization success rate is 16.2\%, comparing to \FileReproducibility success rate for \Tool achieved with up to 16 iterative repair rounds.
The absolute count of \xEFR \nbs peaks at 2 fixes, while \xEFR \nbs become a larger \emph{share} of reproducible outcomes from 3--4 fixes onward (and dominate at higher fix counts despite smaller totals).
Eliminating all errors thus often requires additional fixes beyond the first reproducibility milestone.}

Figure~\ref{fig:fix_boxplot_file} compares the distribution of fix counts by outcome: ``To All'' (all \srep \nbs), ``To E-R'' (\xER), and ``To EF-R'' (\xEFR).
In file-level upgrade over \NumNotebookSubject \nbs, \FileReproducibleNumNotebook \nbs become \reproducible; 
the mean number of fixes per \nb across all upgraded \nbs is \AvgFixIterations, and among \reproducible \nbs the mean is \revise{3.87}.
Among \reproducible \nbs, the mean number of fixes that achieve \reproducibility but leave error cells (\xER) is \revise{2.56}, and the mean number that achieve \reproducibility without error cells (\xEFR) is \revise{4.35}.
``To EF-R'' has a higher median and mean (and wider spread up to 16 fixes), indicating that achieving error-free \reproducibility typically requires more fixes than achieving \reproducibility with residual errors.
Most \nbs are made \reproducible with few fixes: at least half require no more than \revise{2} fixes, while the higher mean of \revise{3.87} reflects the right-skewed tail; achieving \xEFR often requires more fixes than \xER.

To characterize the kind of fixes the LLM applies and the resulting outcomes, Table~\ref{tab:file_state_change} reports state-change transitions by fix type.
Rows identify the fix types selected for the next iteration: the LLM chooses \RuntimeReduction when the script exceeds the runtime limit, \ErrorRepair when the script has execution errors, and \ScoreCalibration when the script has no errors and does not need runtime reduction, but its output must be aligned toward the reported \Kaggle score.
Columns correspond to the status of the \nb after the fix is applied and the \nb is run: Timeout (execution exceeds the time limit), Error Non-reproducible (\xENR), Error-free Non-reproducible (\xEFNR), Error Reproducible (\xER), and Error-free Reproducible (\xEFR).
\ScoreCalibration is the most frequent fix type (\revise{39,226} applications), and the vast majority of those applications leave the \nb \xEFNR (\revise{35,339}); \revise{1,261} reach \xEFR.
\ErrorRepair is the next most frequent (\revise{27,765}); \revise{1,059} reach \xEFR and \revise{834} \xER, but \revise{17,046} remain \xENR and \revise{5,176} become \xEFNR, so error repair often improves state but does not always achieve reproducibility in one step.
\RuntimeReduction is applied \revise{15,718} times; \revise{11,838} applications still result in a timeout, and only \revise{127} reach a \reproducible state (\revise{41} \xER, \revise{86} \xEFR), so reducing runtime is the hardest type of fix to resolve.
Overall, the table shows that \ScoreCalibration and \ErrorRepair account for most fixes.
\ErrorRepair produces the largest number of error-containing \reproducible outcomes, whereas \ScoreCalibration contributes the most \xEFR outcomes. By contrast, \RuntimeReduction rarely leads directly to \reproducibility.




\conclusionbox{
\noindent\textbf{\underline{Summary}:}
\revise{In file-level upgrade (GPT-5.2), \Tool makes \FileReproducibleNumNotebook of \NumNotebookSubject \nbs (\FileReproducibility) \reproducible; a single-fix zero-shot repair succeeds on only 16.2\% of the cohort, versus \FileReproducibility with up to 16 iterative fixes.
Among \reproducible \nbs, the mean is 3.87 LLM fixes (median 2); achieving \xEFR requires more fixes on average (4.35) than \xER (2.56).
By fix type, \ScoreCalibration and \ErrorRepair dominate; \ScoreCalibration contributes the most \xEFR outcomes (1,261), while \RuntimeReduction rarely yields immediate \reproducibility (127 of 15,718).}
}


\begin{table}[t]
\begin{small}
\centering
\caption{Number of LLM fixes by type (rows) and resulting notebook \reproducibility outcomes (columns).}
\label{tab:file_state_change}
\vspace{-5pt}
\resizebox{\linewidth}{!}{
\begin{tabular}{lcccccc}
\hline
\rowcolor[HTML]{E0E0E0}
Fix Type &
  \multicolumn{1}{|l|}{
    \cellcolor[HTML]{E0E0E0}\textbf{\#Total fixes}
  } &
  \textbf{Timeout} &
  \textbf{\begin{tabular}[c]{@{}c@{}}
    Error\\Non-reproducible
  \end{tabular}} &
  \textbf{\begin{tabular}[c]{@{}c@{}}
    Error-free\\Non-reproducible
  \end{tabular}} &
  \textbf{\begin{tabular}[c]{@{}c@{}}
    Error\\Reproducible
  \end{tabular}} &
  \textbf{\begin{tabular}[c]{@{}c@{}}
    Error-free\\Reproducible
  \end{tabular}}
\\ \hline

\textbf{\RuntimeReduction} &
  \multicolumn{1}{|c|}{\revise{15,718}} &
  \revise{11,838} &
  \revise{2,797} &
  \revise{956} &
  \revise{41} &
  \revise{86}
\\

\rowcolor[HTML]{E0E0E0}
\textbf{\ErrorRepair} &
  \multicolumn{1}{|c|}{
    \cellcolor[HTML]{E0E0E0}\revise{27,765}
  } &
  \revise{3,650} &
  \revise{17,046} &
  \revise{5,176} &
  \revise{834} &
  \revise{1,059}
\\

\textbf{\ScoreCalibration} &
  \multicolumn{1}{|c|}{\revise{39,226}} &
  \revise{1,136} &
  \revise{1,475} &
  \revise{35,339} &
  \revise{15} &
  \revise{1,261}
\\ \hline
\end{tabular}
}
\end{small}
\end{table}

\subsection{\UseMacro{rq-similarity}: Code Modification Scale}
\label{sec:eval:rq-similarity}
\begin{figure}[t]
    \centering
    \begin{minipage}[t]{0.46\linewidth}
        \centering
        \includegraphics[width=\linewidth]{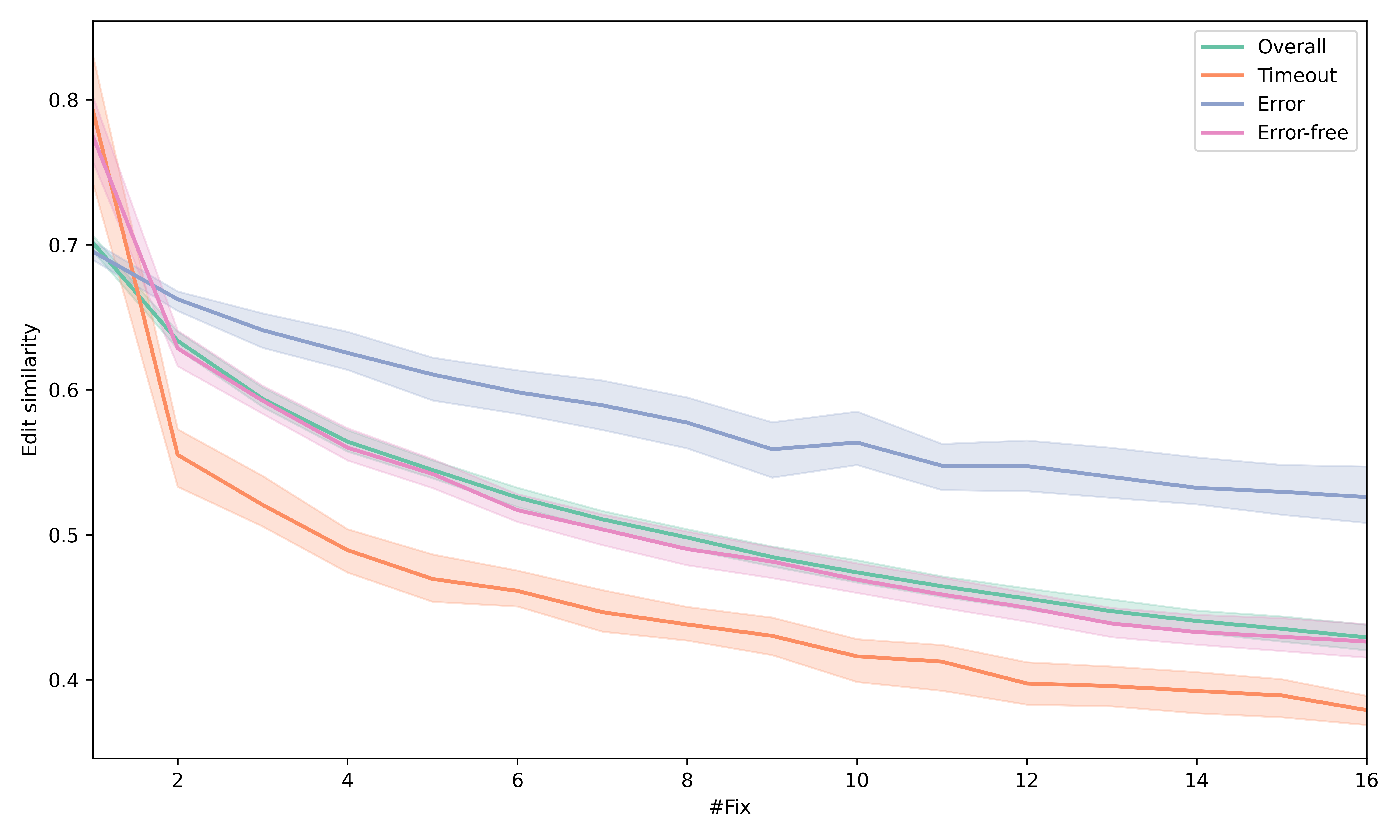}
        \subcaption{\revise{Edit similarity to the original \nb as \#fix increases.}}
        \label{fig:edit_sim_file}
    \end{minipage}
    \hfill
    \begin{minipage}[t]{0.46\linewidth}
        \centering
        \includegraphics[width=\linewidth]{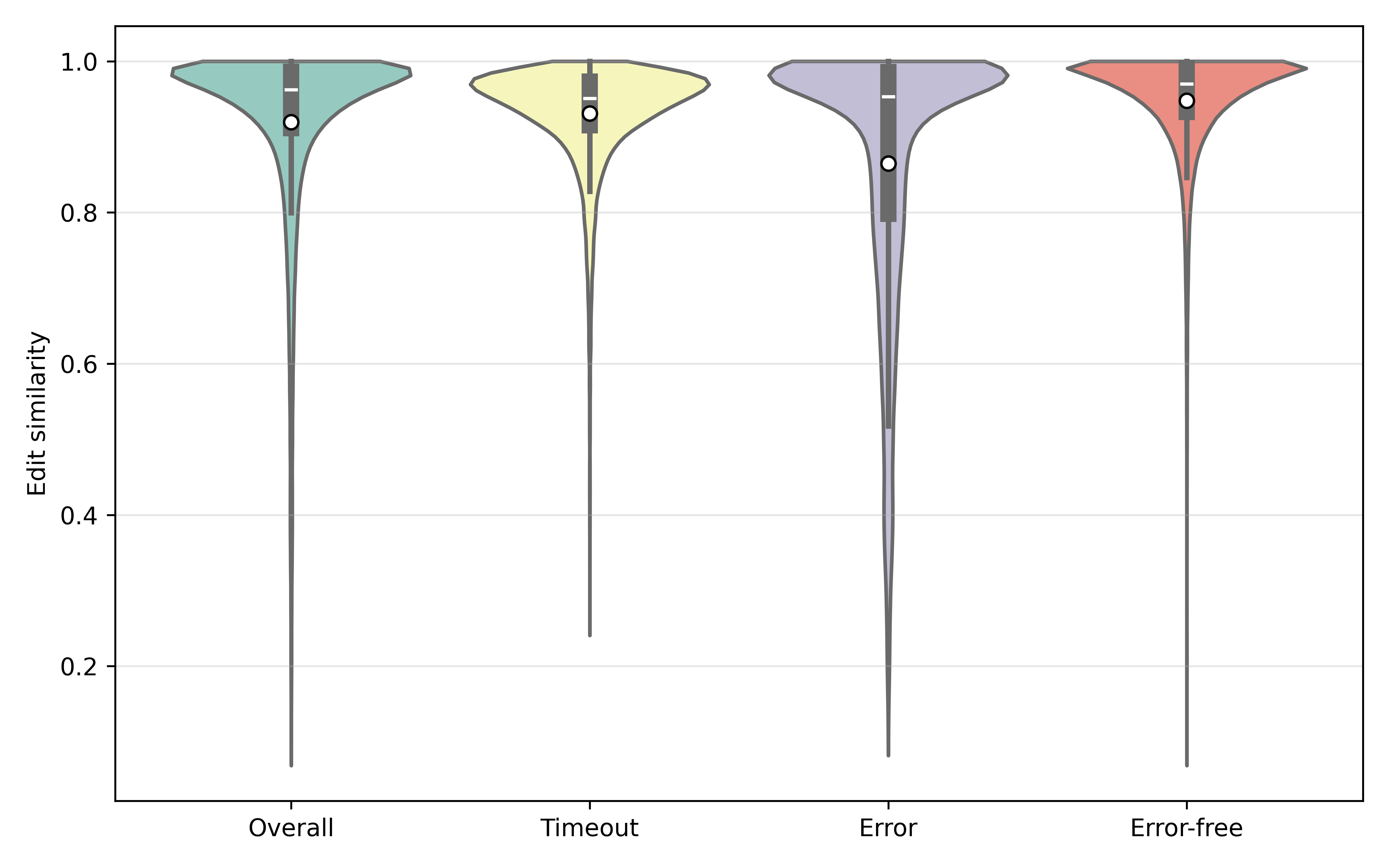}
        \subcaption{\revise{Distribution of the edit similarity between the \nb before and after each LLM fix.}}
        \label{fig:edit_sim_violin}
    \end{minipage}
    \vspace{-8pt}
    \caption{Cumulative (left) and per-fix (right) code modification scale by \Tool \revise{(GPT-5.2)}.}
\end{figure}

Making \nbs \reproducible often requires substantial cumulative code modification. 
Figure~\ref{fig:edit_sim_file} displays how much the \nb changes as \#fix grows at the file level: edit similarity is measured between the current version and the \emph{baseline} version.
Edit similarity to the baseline drops as the number of fixes grows (per file, from \revise{$\sim$0.7 to $\sim$0.43} by 16 fixes).
Per fix, however, the LLM usually makes small, incremental edits (median edit similarity to the previous version is close to 1.0); a minority of fixes involve large changes.
Specifically, four series (Overall, Timeout, Error, Error-free) all trend downward: more fixes lead to greater divergence from the baseline.
The Timeout series has the lowest similarity from the second fix onwards (down to \revise{$\sim$0.38} by 16 fixes), so \nbs that initially time out undergo the largest cumulative modification; the Error series remains higher but still declines to $\sim$0.53.
Thus, we argue that achieving reproducibility requires more than minor tweaks---cumulative changes can substantially alter the code; the extent of modification depends on the initial problem (timeout \nbs change the most).

The LLM primarily performs incremental edits per step, but cumulative file-level divergence from the baseline is substantial when many fixes are applied.
Figure~\ref{fig:edit_sim_violin} reveals edit similarity per fix, comparing each version to the previous version (one fix step).
Four series (Overall, Timeout, Error, Error-free) are concentrated at high similarity (median near 1.0), implying that most individual fixes are small and incremental.
All categories exhibit a long tail to low similarity, so some fixes involve large edits; the overall shape is similar across categories, meaning that the amount of modification per fix does not differ drastically by fix type.

\conclusionbox{
\noindent\textbf{\underline{Summary}:}
Making \nbs \reproducible often requires substantial cumulative code modification: edit similarity to the baseline drops as \#fix grows (from $\sim$0.7 to \revise{$\sim$0.43} by 16 fixes), and \nbs that initially time out change the most. Per fix, the LLM usually makes small, incremental edits (median edit similarity to the previous version close to 1.0); a minority of fixes involve large edits. The LLM thus performs incremental edits per step, but cumulative divergence from the baseline becomes substantial when many fixes are applied.
}

\subsection{\UseMacro{rq-cost}: Cost}
\label{sec:eval:rq-cost}
\revise{With GPT-5.2, applying \Tool remains practical at scale: Table~\ref{tab:file_token} reports the cost (USD) and average token counts at the per-notebook and per-fix level.}
Per notebook, the average cost is \revise{\$0.65}, with average token usage of \revise{52,198} input (not cached), \revise{35,818} input (cached), and \revise{39,119} output.
At the per-fix level, \ErrorRepair costs \revise{\$0.061} on average (\revise{5,507} input not cached, \revise{3,244} cached, \revise{3,642} output), \ScoreCalibration \revise{\$0.061} (\revise{4,928} not cached, \revise{3,822} cached, \revise{3,703} output), and \RuntimeReduction \revise{\$0.059} (\revise{3,829} not cached, \revise{2,474} cached, \revise{3,704} output); \revise{the three fix types have similar per-fix cost, with \RuntimeReduction using fewer cached tokens.}
\revise{GPT-OSS-120b costs are estimated using OpenRouter rates in the same table: only \$0.0074 per notebook on average, with higher uncached input volume ($\sim$64K) but minimal prompt caching ($\sim$1.1K cached).
Although GPT-OSS is less functionally consistent than GPT-5.2 in RQ1, it offers a much cheaper alternative when cost is the primary concern, and some loss in modernization consistency is acceptable.}

\begin{table}[t]
\begin{small}
\centering
\caption{Cost and token usage by \Tool.}
\label{tab:file_token}
\vspace{-6pt}
\subfloat[GPT-5.2]{
\begin{tabular}{ll|c|ccc}
\cline{3-6}
\multicolumn{2}{l|}{\multirow{2}{*}{}} & \multirow{2}{*}{\begin{tabular}[c]{@{}c@{}}Cost\\ (USD)\end{tabular}} & \multicolumn{3}{c}{Average \#Tokens} \\ \cline{4-6} 
\multicolumn{2}{l|}{}                                      &        & Input (not cached) & Input cached & Output    \\ \hline
\multicolumn{2}{l|}{Per \nb}                               & \revise{0.6453} & \revise{52,197.54} & \revise{35,817.71} & \revise{39,118.81} \\ \hline
\multicolumn{1}{l|}{\multirow{3}{*}{Per Fix}} & \ErrorRepair      & \revise{0.0612} & \revise{5,507.29}  & \revise{3,244.12}  & \revise{3,641.50}  \\
\multicolumn{1}{l|}{}                         & \ScoreCalibration & \revise{0.0611} & \revise{4,928.20}  & \revise{3,821.76}  & \revise{3,703.09}  \\
\multicolumn{1}{l|}{}                         & \RuntimeReduction & \revise{0.0590} & \revise{3,829.15}  & \revise{2,474.31}  & \revise{3,703.73}  \\ \hline
\end{tabular}
}
\vspace{5pt}
\subfloat[\revise{GPT-OSS-120b}]{
\revise{\begin{tabular}{ll|c|ccc}
\cline{3-6}
\multicolumn{2}{l|}{\multirow{2}{*}{}} & \multirow{2}{*}{\begin{tabular}[c]{@{}c@{}}Cost\\ (USD)\end{tabular}} & \multicolumn{3}{c}{Average \#Tokens} \\ \cline{4-6} 
\multicolumn{2}{l|}{}                                      &        & Input (not cached) & Input cached & Output    \\ \hline
\multicolumn{2}{l|}{Per \nb}                               & 0.0074 & 63,697.77 & 1,135.59 & 26,962.14 \\ \hline
\multicolumn{1}{l|}{\multirow{3}{*}{Per Fix}} & \ErrorRepair      & 0.0008 & 6,952.89  & 100.78  & 2,763.56  \\
\multicolumn{1}{l|}{}                         & \ScoreCalibration & 0.0007 & 6,132.29  & 146.10  & 2,384.63  \\
\multicolumn{1}{l|}{}                         & \RuntimeReduction & 0.0007 & 4,624.97  & 28.12   & 2,847.35  \\ \hline
\end{tabular}}
}
\end{small}
\end{table}


\begin{figure*}
    \centering
    \begin{minipage}[t]{0.49\linewidth}
        \centering
        \includegraphics[width=\linewidth]{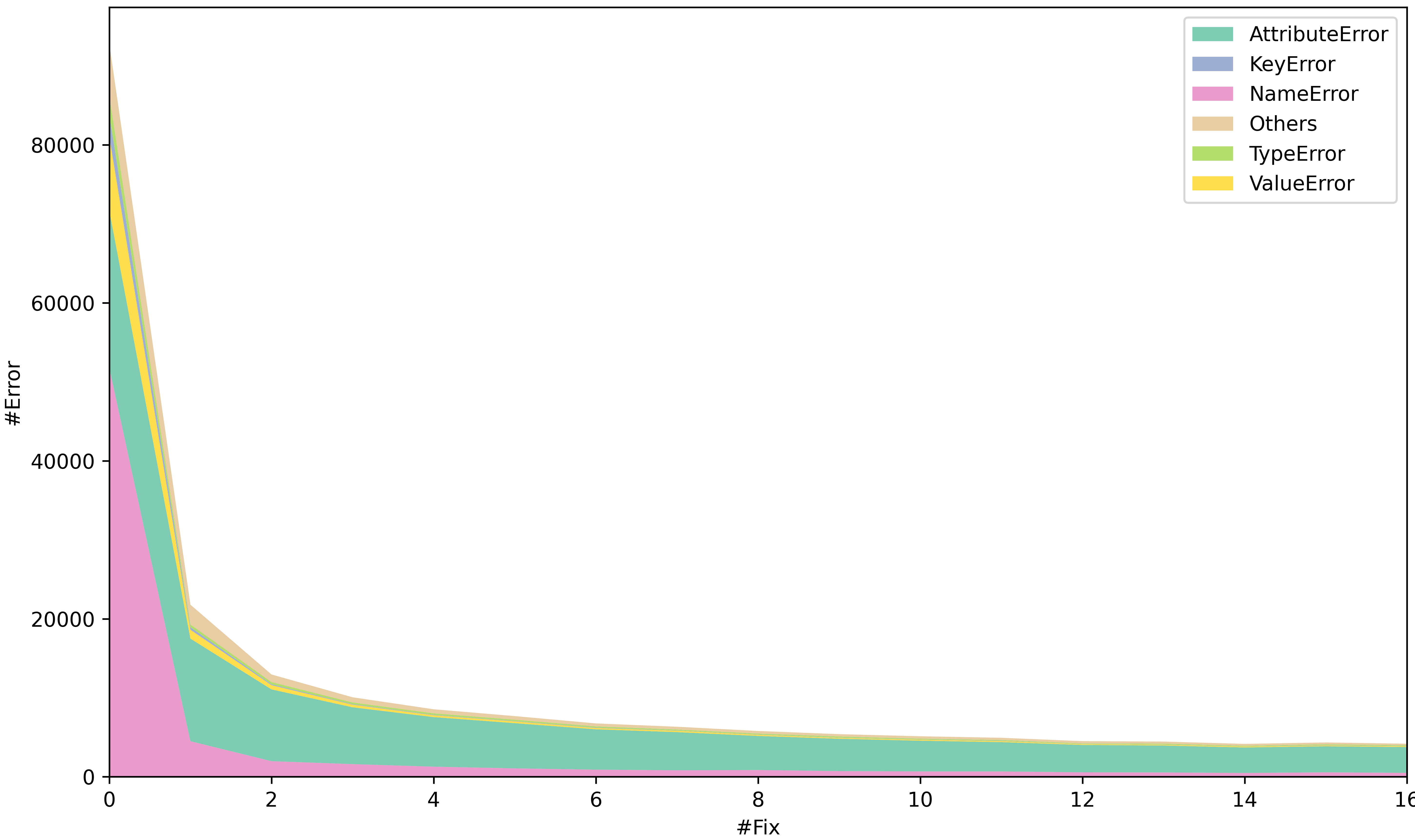}
        \subcaption{\revise{Error types by number of LLM fixes (GPT-5.2).}}
        \label{fig:error_stackplot}
    \end{minipage}
    \hfill
    \begin{minipage}[t]{0.49\linewidth}
        \centering
        \includegraphics[width=\linewidth]{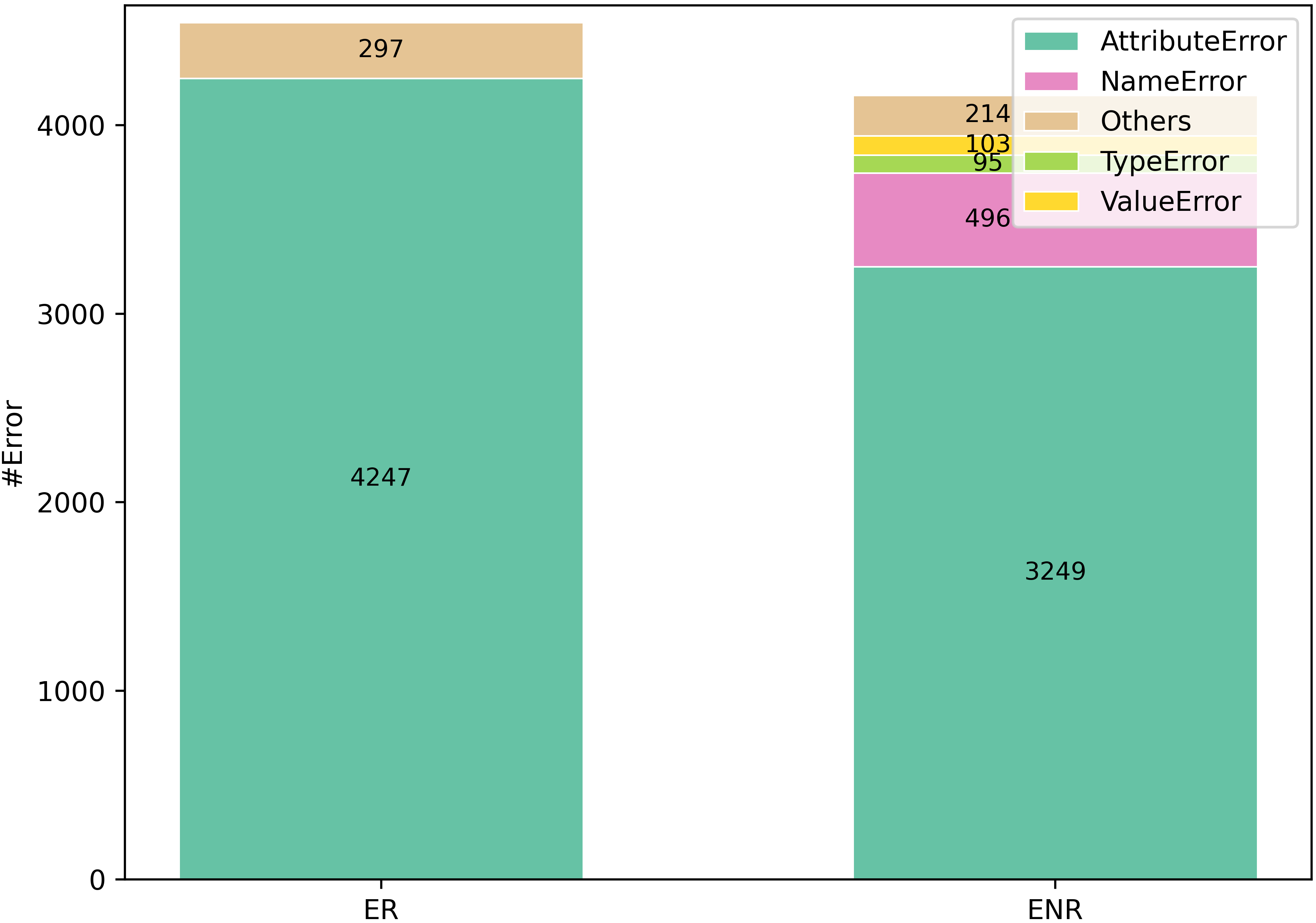}
        \subcaption{\revise{Error types at end of modernization (GPT-5.2).}} 
        \label{fig:error_bar}
    \end{minipage}
    
    \vspace{5pt} 
    
    \begin{minipage}[t]{0.49\linewidth}
        \centering
        \includegraphics[width=\linewidth]{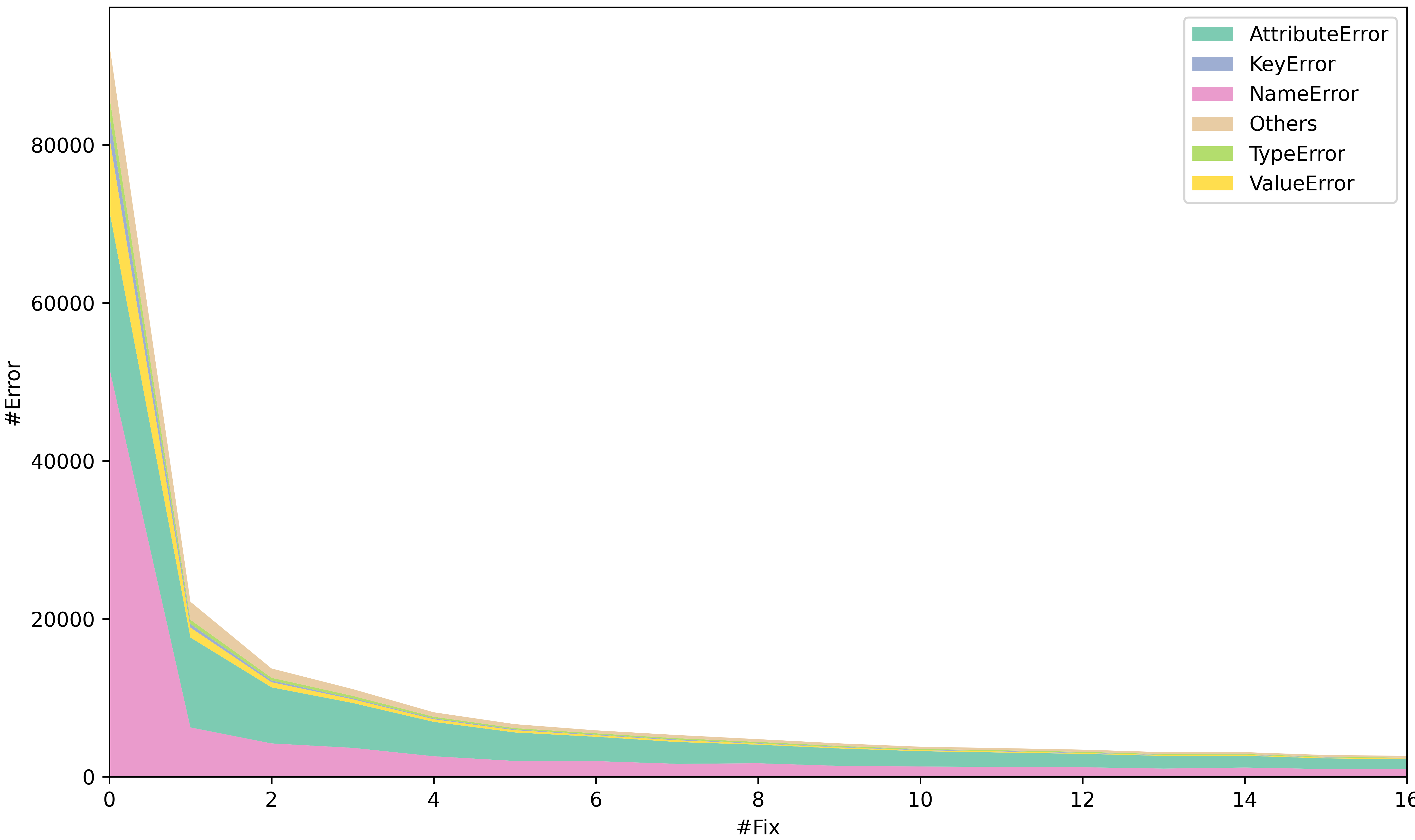}
        \subcaption{\revise{Error types by number of LLM fixes (GPT-OSS).}}
        \label{fig:error_stackplot_oss}
    \end{minipage}
    \hfill
    \begin{minipage}[t]{0.49\linewidth}
        \centering
        \includegraphics[width=\linewidth]{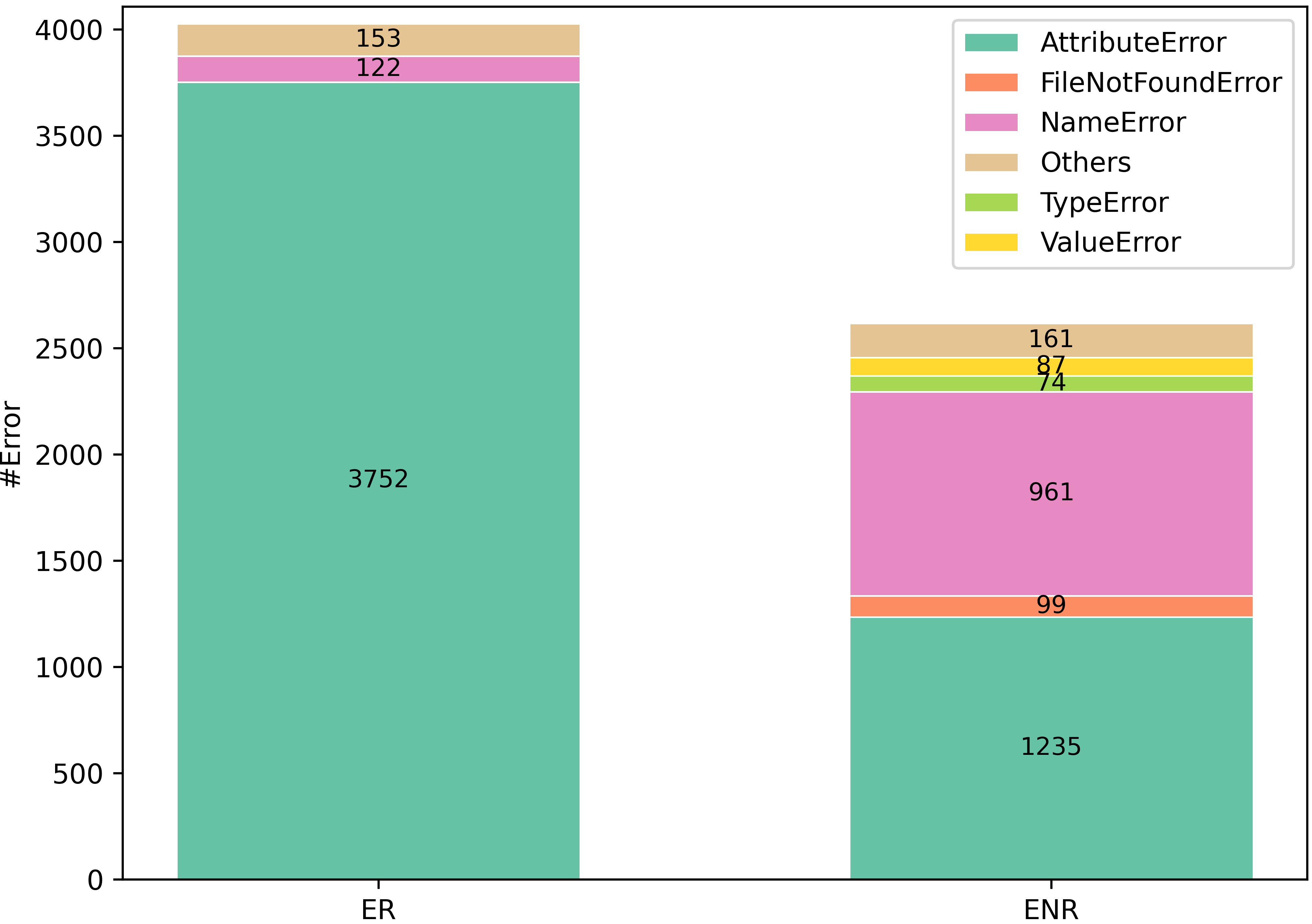}
        \subcaption{\revise{Error types at end of modernization (GPT-OSS).}}
        \label{fig:error_bar_oss}
    \end{minipage}
    
    \vspace{-5pt}
    \caption{Error-type evolution (left) and distribution at end of fixes (right) for both models.}
    \label{fig:overall_error_analysis}
\end{figure*}

\conclusionbox{
\noindent\textbf{\underline{Summary}:}
\revise{With GPT-5.2, upgrading one \nb costs \$0.65 on average ($\sim$52K uncached input, $\sim$36K cached, $\sim$39K output); each fix costs $\sim$\$0.06.
GPT-OSS-120b (OpenRouter pricing) averages \$0.0074 per notebook with far less caching but lower functional consistency, offering a low-cost alternative when budget is the primary constraint.}
}

\subsection{\UseMacro{rq-error-type}: Error Types}
\label{sec:eval:rq-error-type}
We analyze error types in each fix (including \xER \nbs) following the approach from prior work~\cite{WangETAL25NotebookCrash}.
In Python, \emph{NameError} is raised when a name is not defined (\eg \texttt{print(x)} with \texttt{x} undefined or a missing import); \emph{AttributeError} when an object has no requested attribute (\eg \texttt{obj.foo} when \texttt{foo} does not exist); \emph{ValueError} when an operation receives an invalid value of the expected type (\eg \texttt{int('hello')}); \emph{KeyError} when a dictionary key is missing (\eg \texttt{d['key']}); \emph{TypeError} when an operation is applied to an inappropriate type (\eg \texttt{len(5)}).
Such errors often arise from API or dependency changes, incorrect variable scope, or data shape mismatches.

\revise{With GPT-5.2, we find that \textbf{\Tool is highly effective at fixing \emph{NameError}} (undefined name or missing import; baseline 51,772, reduced to 498 after 16 fixes) and substantially reduces \emph{KeyError} and \emph{ValueError}; \emph{AttributeError} (missing or invalid attribute on an object) is the most stubborn---it remains dominant among persistent errors in both \xER and \xENR \nbs after modernization.}
Figure~\ref{fig:error_stackplot} shows how error counts by type evolve as \#fix increases.
At baseline (\#fix\,=\,0), the top five types are \emph{NameError} (51,772), \emph{AttributeError} (19,923), \emph{ValueError} (9,316), \emph{KeyError} (2,773), and \emph{TypeError} (2,407).
\revise{Total errors drop sharply in the first few fixes (from $>$92K to $\sim$20K by \#fix\,=\,1, $\sim$10K by \#fix\,=\,2) and then decline more gradually. 
After 16 fixes, 4,183 crashes remain;
\emph{NameError} (51,772 \(\rightarrow\) 498) falls sharply in early fixes but is not fully eliminated.  \emph{AttributeError} (19,923 \(\rightarrow\) 3,265) becomes the dominant remaining type at higher \#fix.
\Tool resolves many \emph{NameErrors} effectively, but \emph{AttributeErrors} are harder to fix.}
We argue that missing-import or undefined-name issues are amenable to automated repair; attribute and API-usage issues often require deeper or multi-step fixes.

Figure~\ref{fig:error_bar} shows the distribution of error types at each \nb's final fix for \xER and \xENR \nbs (only types $\geq$2\% shown).
\revise{\emph{AttributeError} is the main stubborn type in both outcomes; \emph{NameError} persists chiefly in \xENR \nbs, so fixing \emph{NameError} is often necessary to reach reproducibility, while \emph{AttributeErrors} frequently remain even when the \nb is already \reproducible.}
\revise{In \xER \nbs, \emph{AttributeError} dominates (4,247, 93.5\%), followed by Others (297, 6.5\%). 
}
\revise{In \xENR \nbs, \emph{AttributeError} is again largest (3,249, 78.2\%), followed by \emph{NameError} (496, 11.9\%).}

\revise{Figures~\ref{fig:error_stackplot_oss} and~\ref{fig:error_bar_oss} show the same analysis for GPT-OSS-120b.
After 16 fixes, 2,652 crashes remain---fewer total 
persistent crashes than GPT-5.2 but with a larger residual \emph{NameError} count (498 \(\rightarrow\) 965).
In \xER \nbs, \emph{AttributeError} overwhelmingly dominates (3,752, 93.2\%).
In \xENR \nbs, \emph{AttributeError} (1,235, 47.2\%) and \emph{NameError} (961, 36.7\%) remain the two most prevalent error types.
GPT-OSS leaves substantially more \emph{NameError} in \xENR \nbs than GPT-5.2 (36.7\% vs.\ 11.9\%), consistent with its lower functional-equivalence rate in RQ1.}

\revise{Complementing the automatic error-type taxonomy, we manually label \emph{crash causes} and \emph{crash phases} on the inspected statistically sampled \{before, after\} \nb pairs.
For GPT-5.2, \emph{library issues} (LIB) constitute the most frequent crash root cause (55\% before, 88\% after), followed by \emph{environment issues} (ENV; 39\% before, 2\% after); IRR=0.90.
By crash phase, failures concentrate at \emph{environment setup} (ENVS; 57\% before, 85\% after), then \emph{data preparation} (DATAP; 22\% before, 5\% after); IRR=0.90.
Thus, modernization primarily removes ENV-related failures and DATAP-stage crashes, while residual sampled crashes are increasingly LIB-dominated and ENVS-localized---consistent with persistent \emph{AttributeError}s that reflect library/API drift under the new stack.}

\revise{For GPT-OSS-120b, LIB again dominates (56\% before, 74\% after), but ENV remains more visible after upgrade (37\% before, 9\% after) than under GPT-5.2; IRR=0.84.
By crash phase, ENVS is still primary (57\% before, 73\% after), while DATAP retains a larger post-upgrade share (23\% before, 14\% after) than GPT-5.2 (5\%), indicating that \textbf{the open model less consistently clears early pipeline/setup failures}; IRR=0.91.
Overall, the manual crash analysis shows that \Tool fixes environment and data-preparation blockers first; the hardest remaining failures are library- and setup-stage issues, which align with stubborn \emph{AttributeError}s and with GPT-OSS's higher residual \emph{NameError} burden in \xENR \nbs.}

\conclusionbox{
\noindent\textbf{\underline{Summary}:}
\revise{\Tool substantially reduces crash counts from $>$92K at baseline; after 16 fixes, GPT-5.2 leaves 4,183 persistent crashes and GPT-OSS-120b leaves 2,652.
Both backends eliminate most \emph{NameError}/\emph{KeyError}/\emph{ValueError} early, but \emph{AttributeError} dominates remaining errors in \xER ($>$93\%) and \xENR ($>$78\% for GPT-5.2).
Manual crash-cause/phase labeling shows modernization removes most ENV and DATAP failures; residual crashes are mainly LIB-related at ENVS.
GPT-OSS retains more \emph{NameError} in \xENR and more post-upgrade ENV/DATAP crashes than GPT-5.2, consistent with its weaker functional consistency in RQ1.}
}

\section{Discussion}
\label{sec:discussions}

\subsection{\revise{Alternative Repair Strategy}: \Tool with Cell-Level \ErrorRepair}
\label{sec:discussions:cell-level}


\Tool performs file-level \errorrepair, i.e., attempting to fix all errors at once.
\revise{As a preliminary comparison on a subset of our dataset, we also explore an alternative strategy: cell-level \errorrepair, i.e., attempting to fix one error at a time.}
Specifically, in the \errorrepair prompt, we only provide the \nb cells up to the first cell with error(s), plus its traceback and the following cell, and instruct the LLM to focus on fixing the error.
\revise{Because cell-level repair only diverges from file-level \Tool when errors persist, we compare both repair strategies on the same subset: 2,842 \xENR \nbs that fail on baseline evaluation in a previous version of our experiment.}


\begin{figure}[t]
    \centering
    \begin{minipage}[b]{0.56\linewidth}
        \centering
        \includegraphics[width=\linewidth]{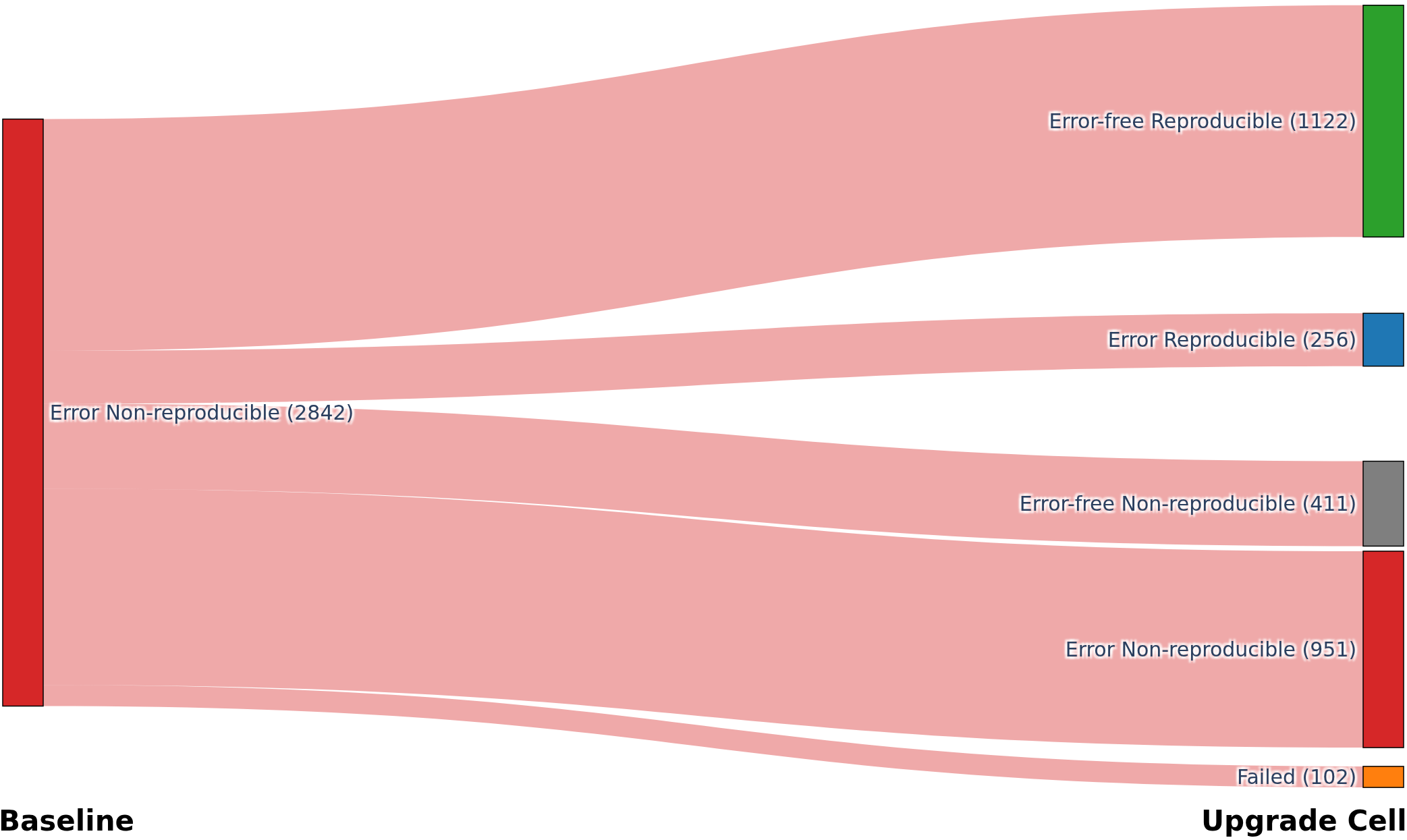}
        \vspace{-15pt}
        \subcaption{\revise{\Nb \reproducibility transitions from \bsl to cell-level \errorrepair.}}
        \label{fig:sankey_baseline_cell}
    \end{minipage}%
    \hfill
    \begin{minipage}[b]{0.38\linewidth}
        \centering
        \includegraphics[width=\linewidth]{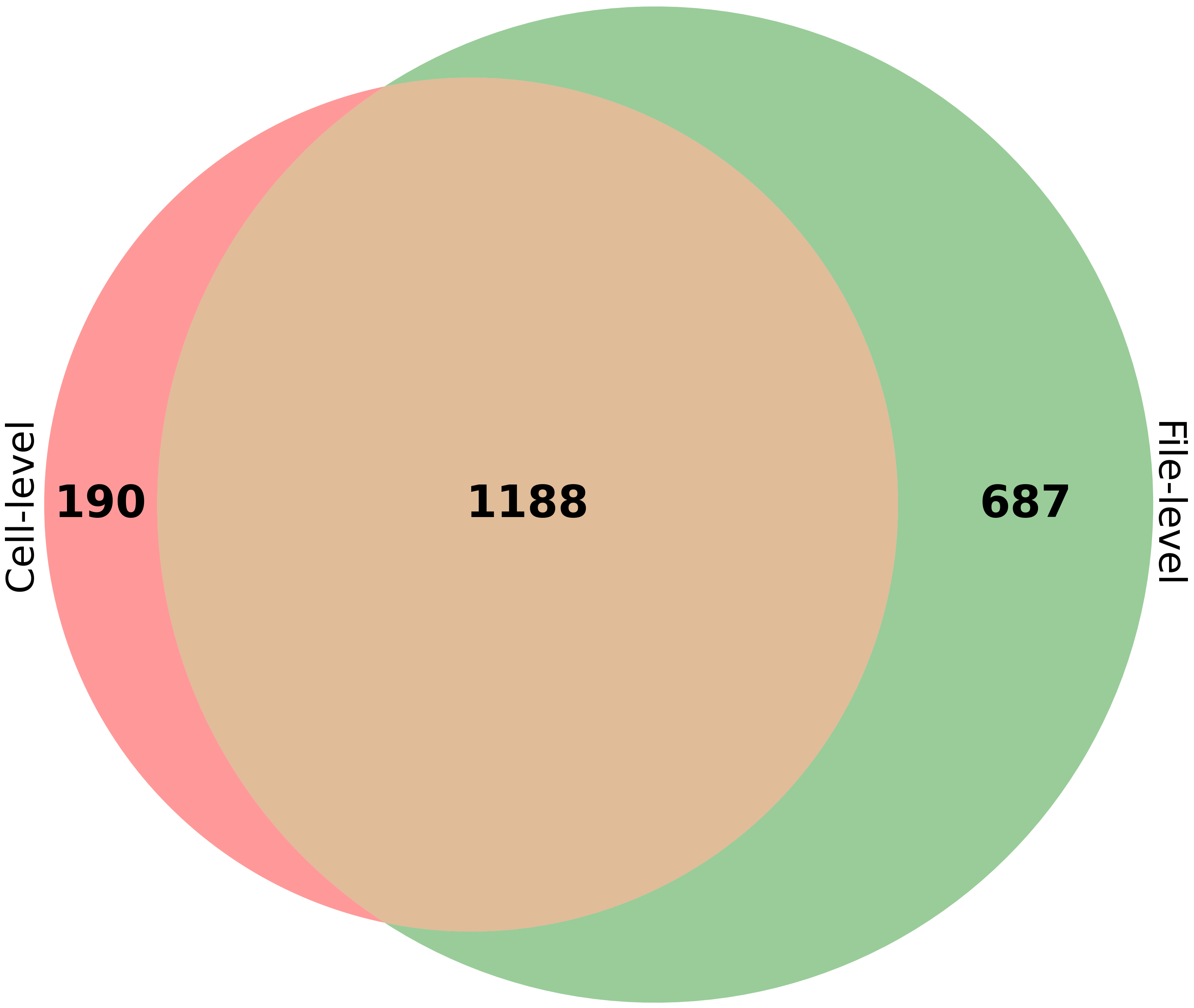}
        \subcaption{\revise{Successfully modernized notebooks by cell- and file-level upgrades.}}
        \label{fig:venn_diagram_cell_vs_file}
    \end{minipage}
    \caption{Effectiveness of using cell-level \errorrepair in \Tool.}
    \label{fig:combined_cell_vs_file}
\end{figure}

Figure~\ref{fig:sankey_baseline_cell} shows the results of cell-level \errorrepair.
With cell-level repair, 1,378 \nbs become \reproducible, while 1,362 remain \nonreproducible, and 102 fail due to timeout/LLM failure.
Figure~\ref{fig:venn_diagram_cell_vs_file} compares which \nbs can be modernized by file-level and cell-level repair.
We find that file-level \errorrepair is slightly more effective, uniquely modernizing 687 notebooks versus 190 for cell-level \errorrepair. 
\revise{Given the stronger coverage of file-level \errorrepair, we therefore focus on file-level repair in this work.}







\subsection{Limitations}
\label{sec:discussions:limitations}

%

\paragraph{Offline Grading and Reproducibility Criteria} Since most \Kaggle competitions do not publicly release their test sets, we use the offline-grading setup from \MLEBench~\cite{Chan24MLEBench} to approximate the \Kaggle evaluation.
This leads to a mismatch between the training and test sets.
\revise{To mitigate this, we classify \reproducibility using a one-sample $t$-test (\S\ref{sec:motivation:baseline:env}).}
Future work can explore other sources of \nbs where the test set is publicly available to enable a stricter \reproducibility criterion.
\paragraph{\revise{Stochasticity of LLM Modernization}}
\revise{Each file-level modernization trajectory is executed once per \nb because re-running thousands of agentic LLM repair loops is prohibitively expensive (total cost ${>}$ \$5k USD).
For \reproducibility \emph{measurement}, we repeat \nb execution up to 10 times and apply one-sample $t$-tests to account for score variance from stochastic training.}
\paragraph{\revise{Generalizability}}
\revise{Our evaluation demonstrates generalizability to both a proprietary LLM (GPT-5.2) and an open-weight LLM (GPT-OSS-120b).
We also observe that successful modernizations share many recurring library/API evolution patterns.
Future work can study extracting generalizable patterns in the form of deterministic refactoring recipes, e.g., OpenRewrite scripts~\cite{OpenRewrite}.}
\paragraph{Flexible Error-Repair and Semantic Preservation.} Some errors do not need to be fixed for \nbs to produce valid and \reproducible predictions. As observed in \S\ref{sec:eval:rq-reproducibility}, around 1/4 of the \reproducible \nbs (GPT-5.2) still contain errors. \new{Additionally, score-based \reproducibility does not guarantee functional equivalence but is acceptable, as \Tool instructs the model to preserve the original semantics while also enabling score calibration. As discussed in \S\ref{sec:eval:rq-reproducibility}, these objectives may cause confusion or lead to repairs that restore the target score at the expense of semantic consistency}.
\new{Future work will prioritize functionality-critical errors, separate structure-preserving repair from score calibration, and strengthen semantic validation to distinguish reproducibility from functional equivalence.}
\section{Related Work}
\label{sec:related}


\paragraph{Reproducibility of ML research and \nbs}
\citet{PimentelETAL19JupyterRepro} study the reproducibility of Jupyter notebooks on GitHub; their follow-up work introduces Julynter~\cite{PimentelETAL21Julynter} for identifying potential reproducibility issues in Jupyter notebooks.
\citet{PineauETAL21ReproducibilityNeurIPS} analyze reproducibility in ML research papers through the NeurIPS reproducibility program, and \citet{GundersenKjensmo18ReproAI,HendersonETAL18DeepRL} study reproducibility challenges in AI and deep RL.
\citet{WangETAL25NotebookCrash} characterize notebook crash types by analyzing error outputs from notebooks mined from GitHub and \Kaggle, without re-executing the notebooks.
\revise{\citet{YaoETAL26NBTest} introduce NBTest, a regression-testing framework that automatically generates cell-level assertions for machine learning notebooks.}
\revise{\citet{elhashemy2025bridging} propose a multi-agent system for transforming notebooks from prototypes to production settings; their focus is workflow migration rather than code modernization under environment erosion.}
In contrast, our study focuses on large-scale re-execution in a containerized environment and evaluates not only execution success but also \scoredeviation against the reported target.

\paragraph{Benchmarks and datasets for MLE \nbs}
\citet{YinETAL23CodeGenNotebooks} build ARCADE, a benchmark for natural-language-to-code generation in data science notebooks, and introduce PACH-INCO, a notebook-centric code LM that reasons over cells and execution context.
\citet{GhahfarokhiETAL24DistilKaggle} release DistilKaggle, a distilled dataset of \Kaggle notebooks and code-quality signals, and \citet{Chan24MLEBench} introduce \MLEBench, an offline benchmark derived from \Kaggle competitions for evaluating ML engineering agents.
\citet{JinFSE25EditNotebooks} study learning to edit ML pipeline code in notebooks using LLMs and find low accuracy even after fine-tuning.
Our work leverages \MLEBench for large-scale, offline grading and studies how to modernize real-world \Kaggle notebooks to restore \reproducibility under contemporary environments.
\citet{DilharaETAL23PyEvolve} mine and transplant evolution patterns in Python ML systems, complementing our code-modernization perspective.

\paragraph{LLM-based program repair and code evolution}
\citet{BouzeniaETAL25RepairAgent} present RepairAgent, an autonomous LLM-based agent for program repair that invokes tools and validation.
\citet{XieETAL25PReMM} propose PReMM, an LLM-based repair technique for multi-method bugs, and \citet{ZhangETAL24PyDex,DeligiannisETAL25RustAssistant} study LLM-driven repair for Python and Rust, respectively.
\citet{WangETAL25DeprecatedAPI} analyze deprecated API usage in LLM-based code completion and propose lightweight mitigation strategies.
Compared with prior program-repair work, our setting emphasizes notebook-specific execution context (cells and state), a fixed modern environment, and \emph{score-aware} repair: a patch is only acceptable if it restores end-to-end execution and brings the \reproducedscore within the \reproducible band of the reported \targetscore.

\section{Conclusions}
\label{sec:conclusions}

We investigate the \reproducibility challenges of MLE \nbs.
In a large-scale study of \NumNotebookAll \Kaggle \nbs, only \BaselineReproducibility are \reproducible, and \envbackporting hurts instead of helping.
We then develop \Tool, an LLM-driven agentic framework for modernizing \nbs toward \reproducibility \new{as an initial exploration of this approach}.
\Tool iteratively executes notebooks and applies three types of targeted fixes: \errorrepair, \runtimereduction, and \scorecalibration.
Our evaluation shows that \Tool successfully modernizes \FileReproducibility--44.9\% of the \nbs that are \nonreproducible in the baseline environment.
These results suggest that, although code modernization is a best-effort recovery technique, it offers a viable path to restoring \reproducibility for legacy MLE notebooks, enabling more reliable validation and reuse of ML pipelines in a rapidly evolving hardware and software environment.

\begin{acks}                            

We thank Ian Chen, Yuntian Deng, Yinxi Li, Yu Liu, Chengnian Sun, Shirley Xiao, and the anonymous reviewers for their comments and feedback.
This work is partially supported by the Natural Sciences and Engineering Research Council of Canada (NSERC) under funding reference RGPIN2024-04909.
Bihui Jin is additionally supported by the NSERC Canada Graduate Research Scholarship--Doctoral Program (CGRS D).

\end{acks}

\clearpage\newpage
\section*{Data Availability}
\revise{We provide an artifact package containing our \Tool implementation, replication scripts, the \MLEBench grader components used for evaluation, and the curated list of notebook identifiers and metadata from Meta Kaggle and Meta Kaggle Code~\cite{MetaKaggle,MetaKaggleCode}.}
\revise{Notebook content and metadata were collected from \Kaggle's official Meta Kaggle data dumps, as described in \S\ref{sec:motivation:baseline:data}.}
We do not redistribute the competition datasets themselves, which can be downloaded on demand using our provided scripts.
Our artifact is available at \ToolURL \xspace.


\bibliography{bib}

@String{AAAI = "AAAI Conference on Artificial Intelligence"}

@String{ACL = "Annual Meeting of the Association for Computational Linguistics"}

@String{ASE = "International Conference on Automated Software Engineering"}

@String{ASEW = ASE # " Workshops"}

@String{CHI = "CHI Conference on Human Factors in Computing Systems"}

@String{EMSE = "Empirical Software Engineering"}

@String{FSECompanion = "Companion Proceedings of the " # FSE}

@String{ICLR = "International Conference on Learning Representations"}

@String{ICSE = "International Conference on Software Engineering"}

@String{JMLR = "Journal of Machine Learning Research"}

@String{MSR = "International Working Conference on Mining Software Repositories"}

@String{NeurIPS = "Conference on Neural Information Processing Systems"}

@String{PACMPL = "Proceedings of the ACM on Programming Languages"}

@String{TSE = "Transactions on Software Engineering"}

@inproceedings{PimentelETAL19JupyterRepro,
  author = {Pimentel, Jo{\~a}o Felipe and Murta, Leonardo and Braganholo, Vanessa and Freire, Juliana},
  title = {A Large-scale Study about Quality and Reproducibility of {J}upyter Notebooks},
  booktitle = MSR,
  pages = {507--517},
  LONGpublisher = {IEEE},
  LONGaddress = {Piscataway, NJ, USA},
  year = {2019},
  doi = {10.1109/MSR.2019.00077},
}

@article{PimentelETAL21Julynter,
  author = {Pimentel, Jo{\~a}o Felipe and Murta, Leonardo and Braganholo, Vanessa and Freire, Juliana},
  title = {Understanding and Improving the Quality and Reproducibility of {J}upyter Notebooks},
  journal = EMSE,
  volume = {26},
  number = {4},
  articleno = {65},
  numpages = {41},
  year = {2021},
  doi = {10.1007/s10664-021-09961-9},
  url = {https://doi.org/10.1007/s10664-021-09961-9},
}

@article{PineauETAL21ReproducibilityNeurIPS,
  author = {Pineau, Joelle and Vincent{-}Lamarre, Philippe and Sinha, Koustuv and Larivi{\`e}re, Vincent and Beygelzimer, Alina and d'Alch{\'e}{-}Buc, Florence and Fox, Emily and Larochelle, Hugo},
  title = {Improving Reproducibility in Machine Learning Research (A Report from the {NeurIPS} 2019 Reproducibility Program)},
  journal = JMLR,
  volume = {22},
  number = {164},
  pages = {1--20},
  year = {2021},
  url = {https://www.jmlr.org/papers/v22/20-303.html}
}

@inproceedings{GundersenKjensmo18ReproAI,
  author = {Gundersen, Odd Erik and Kjensmo, Sigbj{\o}rn},
  title = {State of the Art: Reproducibility in Artificial Intelligence},
  booktitle = AAAI,
  pages = {1644--1651},
  year = {2018},
  doi = {10.1609/aaai.v32i1.11503},
}

@inproceedings{HendersonETAL18DeepRL,
  author = {Henderson, Peter and Islam, Riashat and Bachman, Philip and Pineau, Joelle and Precup, Doina and Meger, David},
  title = {Deep Reinforcement Learning That Matters},
  booktitle = AAAI,
  pages = {3207--3214},
  year = {2018},
  doi = {10.1609/aaai.v32i1.11694},
}

@incollection{KluyverETAL16Jupyter,
  author = {Kluyver, Thomas and Ragan{-}Kelley, Benjamin and P{\'e}rez, Fernando and Granger, Brian and Bussonnier, Matthias and Frederic, Jonathan and Kelley, Kyle and Hamrick, Jessica and Grout, Jason and Corlay, Sylvain and Ivanov, Paul and Avila, Dami{\'a}n and Abdalla, Safia and Willing, Carol and {{Jupyter Development Team}}},
  title = {Jupyter Notebooks---a publishing format for reproducible computational workflows},
  booktitle = {Positioning and Power in Academic Publishing: Players, Agents and Agendas},
  LONGeditor = {Loizides, Fernando and Schmidt, Birgit},
  pages = {87--90},
  LONGpublisher = {IOS Press},
  LONGaddress = {Amsterdam, The Netherlands},
  year = {2016},
  doi = {10.3233/978-1-61499-649-1-87},
}

@inproceedings{RuleETAL18Notebooks,
  author = {Rule, Adam and Tabard, Aur{\'e}lien and Hollan, James D.},
  title = {Exploration and Explanation in Computational Notebooks},
  booktitle = CHI,
  articleno = {32},
  pages = {1--12},
  numpages = {12},
  LONGpublisher = {Association for Computing Machinery},
  LONGaddress = {New York, NY, USA},
  year = {2018},
  doi = {10.1145/3173574.3173606},
}

@inproceedings{DilharaETAL23PyEvolve,
  author = {Dilhara, Malinda and Dig, Danny and Ketkar, Ameya},
  title = {{PyEvolve}: Automating Frequent Code Changes in {P}ython {ML} Systems},
  booktitle = ICSE,
  pages = {995--1007},
  LONGpublisher = {IEEE},
  LONGaddress = {Piscataway, NJ, USA},
  year = {2023},
  doi = {10.1109/ICSE48619.2023.00091},
}

@article{WangETAL21DocumentationNotebook,
  author = {Wang, April Yi and Wang, Dakuo and Drozdal, Jaimie and Muller, Michael and Park, Soya and Weisz, Justin D. and Liu, Xuye and Wu, Lingfei and Dugan, Casey},
  title = {Documentation Matters: Human-Centered {AI} System to Assist Data Science Code Documentation in Computational Notebooks},
  journal = {Transactions on Computer-Human Interaction},
  volume = {29},
  number = {2},
  articleno = {17},
  pages = {1--33},
  numpages = {33},
  year = {2022},
  doi = {10.1145/3489465},
  url = {https://doi.org/10.1145/3489465},
}

@inproceedings{GhahfarokhiETAL24DistilKaggle,
  author = {Mostafavi Ghahfarokhi, Mojtaba and Asgari, Arash and Abolnejadian, Mohammad and Heydarnoori, Abbas},
  title = {{DistilKaggle}: A Distilled Dataset of {K}aggle {J}upyter Notebooks},
  booktitle = MSR,
  year = {2024},
  pages = {647--651},
  LONGpublisher = {IEEE},
  LONGaddress = {Piscataway, NJ, USA},
  doi = {10.1145/3643991.3644882},
}

@inproceedings{YinETAL23CodeGenNotebooks,
  author = {Yin, Pengcheng and Li, Wen-Ding and Xiao, Kefan and Rao, Abhishek and Wen, Yeming and Shi, Kensen and Howland, Joshua and Bailey, Paige and Catasta, Michele and Michalewski, Henryk and Polozov, Oleksandr and Sutton, Charles},
  title = {Natural Language to Code Generation in Interactive Data Science Notebooks},
  booktitle = ACL,
  year = {2023},
  pages = {126--173},
  LONGpublisher = {Association for Computational Linguistics},
  LONGaddress = {Toronto, Canada},
  doi = {10.18653/v1/2023.acl-long.9},
}

@article{Cohen60kappa,
  author = {Cohen, Jacob},
  title = {A Coefficient of Agreement for Nominal Scales},
  journal = {Educational and Psychological Measurement},
  volume = {20},
  number = {1},
  pages = {37--46},
  year = {1960},
  doi = {10.1177/001316446002000104},
}

@ARTICLE{SeamanTSE99ClosedCodingProcedure,
  author = {Seaman, Carolyn B.},
  journal = TSE,
  title = {Qualitative Methods in Empirical Studies of Software Engineering},
  year = {1999},
  volume = {25},
  number = {4},
  pages = {557--572},
  doi = {10.1109/32.799955},
}

@article{WangETAL25NotebookCrash,
  title={Why do Machine Learning Notebooks Crash? An Empirical Study on Public {P}ython {J}upyter Notebooks},
  author={Wang, Yiran and Meijer, Willem and Hern{\'a}ndez L{\'o}pez, Jos{\'e} Antonio and Nilsson, Ulf and Varr{\'o}, D{\'a}niel},
  journal=TSE,
  year={2025},
  volume={51},
  number={7},
  pages={2181--2196},
  doi={10.1109/TSE.2025.3574500},
}

@inproceedings{YaoETAL26NBTest,
  author = {Yao, Yingao (Elaine) and Nimje, Vedant and Viswanath, Varun and Dutta, Saikat},
  title = {Automated Assertion Generation and Regression Testing for Machine Learning Notebooks},
  booktitle = ASE,
  year = {2026},
  pages = {to appear},
  url = {https://arxiv.org/abs/2509.13656},
}

@inproceedings{JinFSE25EditNotebooks,
author = {Jin, Bihui and Wang, Jiayue and Nie, Pengyu},
title = {Learning to Edit Interactive Machine Learning Notebooks},
year = {2025},
isbn = {9798400712760},
LONGpublisher = {Association for Computing Machinery},
LONGaddress = {New York, NY, USA},
booktitle = FSECompanion,
pages = {681--685},
numpages = {5},
doi = {10.1145/3696630.3728523},
url = {https://doi.org/10.1145/3696630.3728523},
}

@online{OpenRewrite,
  author = {{OpenRewrite}},
  title = {{OpenRewrite by Moderne}: Large Scale Automated Refactoring},
  year = 2026,
  url = {https://docs.openrewrite.org/},
  urldate = {2026-07-26},
}

@inproceedings{elhashemy2025bridging,
  author = {Elhashemy, Hanya and Lotfy, Youssef and Tang, Yongjian},
  title = {Bridging the Prototype-Production Gap: A Multi-Agent System for Notebooks Transformation},
  booktitle = ASEW,
  year = {2025},
  pages = {299--302},
  LONGpublisher = {IEEE},
  LONGaddress = {Piscataway, NJ, USA},
  doi = {10.1109/ASEW67777.2025.00061},
  url = {https://doi.org/10.1109/ASEW67777.2025.00061},
}

@online{MetaKaggle,
  author = {Kaggle},
  title = {Meta Kaggle},
  year = 2025,
  url = {https://www.kaggle.com/datasets/kaggle/meta-kaggle/data},
  urldate = {2025-05-31},
}

@online{MetaKaggleCode,
  author = {Kaggle},
  title = {Meta Kaggle Code},
  year = 2025,
  url = {https://www.kaggle.com/datasets/kaggle/meta-kaggle-code/data},
  urldate = {2025-05-31},
}

@online{pigar,
  author = {{damnever}},
  title = {pigar 2.2.0},
  year = 2025,
  url = {https://pypi.org/project/pigar/2.2.0/},
  urldate = {2025-12-10}
}

@online{KaggleRunSetup,
  author = {Kaggle},
  title = {Notebooks Documentation: The Notebook Environment},
  year = 2025,
  url = {https://www.kaggle.com/docs/notebooks#the-notebooks-environment},
  urldate = {2025-12-10}
}

@online{KaggleTechSpecifications,
  author = {Kaggle},
  title = {Notebooks Documentation: Technical Specifications},
  year = 2025,
  url = {https://www.kaggle.com/docs/notebooks#technical-specifications},
  urldate = {2025-12-10}
}

@online{tiktoken,
  author = {Shantanu Jain},
  title = {tiktoken 0.12.0},
  year = 2025,
  url = {https://pypi.org/project/tiktoken/0.12.0/},
  urldate = {2025-12-10}
}

@inproceedings{Chan24MLEBench,
  title = {{MLE-bench}: Evaluating Machine Learning Agents on Machine Learning Engineering},
  author = {Chan, Jun Shern and Chowdhury, Neil and Jaffe, Oliver and Aung, James and Sherburn, Dane and Mays, Evan and Starace, Giulio and Liu, Kevin and Maksin, Leon and Patwardhan, Tejal and Weng, Lilian and M{\k{a}}dry, Aleksander},
  booktitle = ICLR,
  year = {2025},
  LONGpublisher = {OpenReview.net},
  LONGaddress = {Singapore},
  numpages = {22},
  url = {https://proceedings.iclr.cc/paper_files/paper/2025/hash/7e3767db483c942b883eb4f8cfb74e31-Abstract-Conference.html},
}

@article{ZhangETAL24PyDex,
author = {Zhang, Jialu and Cambronero, Jos\'{e} Pablo and Gulwani, Sumit and Le, Vu and Piskac, Ruzica and Soares, Gustavo and Verbruggen, Gust},
title = {PyDex: Repairing Bugs in Introductory Python Assignments using LLMs},
year = {2024},
issue_date = {April 2024},
LONGpublisher = {Association for Computing Machinery},
LONGaddress = {New York, NY, USA},
volume = {8},
number = {OOPSLA1},
doi = {10.1145/3649850},
journal = PACMPL,
month = apr,
articleno = {133},
pages = {1100--1124},
numpages = {25},
}

@inproceedings{WangETAL25DeprecatedAPI,
author = {Wang, Chong and Huang, Kaifeng and Zhang, Jian and Feng, Yebo and Zhang, Lyuye and Liu, Yang and Peng, Xin},
title = {LLMs Meet Library Evolution: Evaluating Deprecated API Usage in LLM-Based Code Completion},
year = {2025},
isbn = {9798331505691},
LONGpublisher = {IEEE Press},
url = {https://doi.org/10.1109/ICSE55347.2025.00245},
doi = {10.1109/ICSE55347.2025.00245},
booktitle = ICSE,
pages = {885--897},
numpages = {13},
LONGaddress = {Piscataway, NJ, USA},
}

@inproceedings{BouzeniaETAL25RepairAgent,
author = {Bouzenia, Islem and Devanbu, Premkumar and Pradel, Michael},
title = {RepairAgent: An Autonomous, LLM-Based Agent for Program Repair},
year = {2025},
isbn = {9798331505691},
LONGpublisher = {IEEE Press},
LONGaddress = {Piscataway, NJ, USA},
url = {https://doi.org/10.1109/ICSE55347.2025.00157},
doi = {10.1109/ICSE55347.2025.00157},
booktitle = ICSE,
pages = {2188--2200},
numpages = {13},
location = {Ottawa, Ontario, Canada},
series = {ICSE '25}
}

@inproceedings{DeligiannisETAL25RustAssistant,
author = {Deligiannis, Pantazis and Lal, Akash and Mehrotra, Nikita and Poddar, Rishi and Rastogi, Aseem},
title = {RustAssistant: Using LLMs to Fix Compilation Errors in Rust Code},
year = {2025},
isbn = {9798331505691},
LONGpublisher = {IEEE Press},
LONGaddress = {Piscataway, NJ, USA},
url = {https://doi.org/10.1109/ICSE55347.2025.00022},
doi = {10.1109/ICSE55347.2025.00022},
booktitle = ICSE,
pages = {3097--3109},
numpages = {13},
}

@article{XieETAL25PReMM,
author = {Xie, Linna and Li, Zhong and Pei, Yu and Wen, Zhongzhen and Liu, Kui and Zhang, Tian and Li, Xuandong},
title = {PReMM: LLM-Based Program Repair for Multi-method Bugs via Divide and Conquer},
year = {2025},
issue_date = {October 2025},
LONGpublisher = {Association for Computing Machinery},
LONGaddress = {New York, NY, USA},
volume = {9},
number = {OOPSLA2},
url = {https://doi.org/10.1145/3763097},
doi = {10.1145/3763097},
journal = PACMPL,
month = oct,
articleno = {319},
pages = {1316--1344},
numpages = {29},
}



\end{document}